\documentclass[twopage,11pt] {article}  % Specifies the document style.
\usepackage{amssymb}
\usepackage{bm}
\usepackage[dvips]{epsfig}
\usepackage{amssymb}
\input{epsf}
\usepackage{bm}
\usepackage{dcolumn}
\def\rf#1{(\ref{#1})}
\def\lab#1{\label{#1}}
\def\nonu{\nonumber}
\def\br{\begin{eqnarray}}
\def\er{\end{eqnarray}}
\def\be{\begin{equation}}
\def\ee{\end{equation}}

\def\({\left(}
\def\){\right)}

\relax

\def\a{\alpha}

\def\b{\beta}

\def\d{\delta}
\def\D{\Delta}

\def\g{\gamma}
\def\G{\Gamma}

\def\l{\lambda}

\def\n{\nu}

\def\pa{\partial}

\def\ra{\rightarrow}
\def\s{\sigma}

\def\th{\theta}

\def\tp0{\Theta_{+}^{(0)}}
\def\tm0{\Theta_{-}^{(0)}}

\def\vp{\varphi}

\def\f#1#2#3 {f^{#1#2}_{#3}}

\def\win1{{\sf w_{1+\infty}}}

\def\Win1{{\sf W_{1+\infty}}}

\def\rlx{\relax\leavevmode}
\def\inbar{\vrule height1.5ex width.4pt depth0pt}
\def\IZ{\rlx\hbox{\sf Z\kern-.4em Z}}
\def\IR{\rlx\hbox{\rm I\kern-.18em R}}
\def\IC{\rlx\hbox{\,$\inbar\kern-.3em{\rm C}$}}
\def\IN{\rlx\hbox{\rm I\kern-.18em N}}
\def\IO{\rlx\hbox{\,$\inbar\kern-.3em{\rm O}$}}
\def\IP{\rlx\hbox{\rm I\kern-.18em P}}
\def\IQ{\rlx\hbox{\,$\inbar\kern-.3em{\rm Q}$}}
\def\IF{\rlx\hbox{\rm I\kern-.18em F}}
\def\IG{\rlx\hbox{\,$\inbar\kern-.3em{\rm G}$}}
\def\IH{\rlx\hbox{\rm I\kern-.18em H}}
\def\II{\rlx\hbox{\rm I\kern-.18em I}}
\def\IK{\rlx\hbox{\rm I\kern-.18em K}}
\def\IL{\rlx\hbox{\rm I\kern-.18em L}}
\def\one{\hbox{{1}\kern-.25em\hbox{l}}}
\def\0#1{\relax\ifmmode\mathaccent"7017{#1}%
B        \else\accent23#1\relax\fi}

                \def\CSF#1#2#3{{\sl Chaos, Solitons and Fractals} {\bf C#1} (#2) #3}
\def\EPJC#1#2#3{{\sl Eur. Phys. J.} {\bf C#1} (#2) #3}

                \def\JHEP#1#2#3{{\sl JHEP} {\bf#1} (#2) #3}
                \def\PRL#1#2#3{{\sl Phys. Rev. Lett.} {\bf#1} (#2) #3}
                \def\NPB#1#2#3{{\sl Nucl. Phys.} {\bf B#1} (#2) #3}
                
                \def\CMP#1#2#3{{\sl Commun. Math. Phys.} {\bf #1} (#2) #3}
                \def\PRD#1#2#3{{\sl Phys. Rev.} {\bf D#1} (#2) #3}
\def\PRA#1#2#3{{\sl Phys. Rev.} {\bf A#1} (#2) #3}
                
                \def\PLA#1#2#3{{\sl Phys. Lett.} {\bf #1A} (#2) #3}
                \def\PLB#1#2#3{{\sl Phys. Lett.} {\bf #1B} (#2) #3}
                \def\JMP#1#2#3{{\sl J. Math. Phys.} {\bf #1} (#2) #3}

                \def\AoP#1#2#3{{\sl Annals Phys.} {\bf #1} (#2) #3}
                
                \def\RMP#1#2#3{{\sl Rev. Mod. Phys.} {\bf #1} (#2) #3}
                \def\PR#1#2#3{{\sl Phys. Reports} {\bf #1} (#2) #3}

                \def\JPA#1#2#3{{\sl J. Physics} {\bf A#1} (#2) #3}

                \def\PD#1#2#3{{\sl Physica} {\bf D#1} (#2) #3}

\def\EPJC#1#2#3{{\sl Eur. Phys. J.} {\bf C#1} (#2) #3}
\def\JPG#1#2#3{{\sl J. Phys.} {\bf G#1} (#2) #3}

                \def\a{\alpha}
                \def\b{\beta}

                \def\d{\delta}
                \def\D{\Delta}
                
                \def\g{\gamma}
                \def\G{\Gamma}
                \def\vp{\varphi}

                \def\/{\frac}

                \def\l{\lambda}

                \def\n{\nu}

                \def\pa{\partial}

                \def\ra{\rightarrow}
                \def\rh{\rho}
                \def\vp{\varphi}
                
                \def\s{\sigma}
                
                \def\th{\theta}

                \def\({\Big(}
                \def\){\Big)}
                \def\[{\Big[}
                \def\]{\Big]}

\begin{document}           % End of preamble and beginning of text.
\begin{center}
  {\large\bf Solitons as baryons and qualitons as constituent quarks in two-dimensional QCD}
\end{center}

\begin{center}

H. Blas $^{a}$ and H.L. Carrion $^{b}$

\vspace{.5 cm} \small

\par \vskip .1in \noindent

$^{a}$ Departamento de F\'{\i}sica - ICET\\
Universidade Federal de Mato Grosso\\
Av. Fernando Correa, s/n, Coxip\'o \\
78060-900, Cuiab\'a - MT - Brazil\\
$^{b}$ Instituto de F\'{\i}sica, Universidade de S\~ao Paulo,
\\
Caixa Postal 66318, 05315-970, S\~ao Paulo, SP, Brazil.

\end{center}

%\maketitle                 % Produces the title.

% Set to use the "plain" pagestyle
\pagestyle{myheadings} \thispagestyle{plain} \markboth{Harold
Blas}{Baryons and quark solitons in QCD$_{2}$}
\setcounter{page}{1}

\begin{abstract}
We study the soliton type solutions arising in two-dimensional
quantum chromodynamics (QCD$_{2}$). In bosonized QCD$_{2}$ these type of
solutions emerge as describing baryons and quark solitons
(excitations with ``colored'' states), respectively. The so-called
generalized sine-Gordon model (GSG) arises as the low-energy
effective action of bosonized QCD$_{2}$ for unequal quark mass
parameters, and it has been shown that the relevant solitons
describe the normal and exotic baryonic spectrum of QCD$_{2}$
[JHEP(03)(2007)(055)]. In the first part of this chapter we
classify the soliton and kink type solutions of the sl(3) GSG model with
three real fields, which corresponds to QCD$_{2}$ with three flavors. Related to the GSG model we consider the
sl(3) affine Toda model coupled to matter fields (Dirac spinors)
  (ATM). The strong coupling sector is described by the
  $sl(3)$ GSG model which completely decouples from the Dirac spinors. In the
spinor sector we are left with Dirac fields coupled to GSG fields.
Based on the equivalence between the U(1) vector and topological
currents, which holds in the ATM model, it has been shown the
confinement of the spinors inside the solitons and kinks of the
GSG model providing an extended hadron model for ``quark"
confinement [JHEP(01)(2007)(027)]. Moreover, it has been proposed
that the constituent quark in QCD is a topological soliton. These
qualitons (quark solitons), topological excitations with the
quantum numbers of quarks, may provide an accurate description of
what is meant by constituent quarks in QCD. In the second part of
this chapter we discuss the appearance of these type of quark
solitons in the context of bosonized QCD$_{2}$ (with $N_{f}=1$ and
$N_{c}$ colors) and the relevance of the $sl(2)$ ATM model in order to
describe the confinement of the color degrees of freedom. We
have shown that QCD$_{2}$ has quark soliton solutions if the quark
mass is sufficiently large.
\end{abstract}

%\end{titlepage}
                %\noindent PACS numbers: 04.20.Cv, 03.65.Ta, 04.80.Cc

             \section{Introduction}

A useful theoretical laboratory for studying several problems in
Quantum Chromodynamics is QCD in two dimensions \cite{abdalla,
frishman}. This theory can be written in bosonized form
\cite{witten1} for arbitrary numbers of colors $N_{c}$ and flavors
$N_{f}$ \cite{date1}. It reflects accurately the phenomena of
quark confinement and condensation in the vacuum that we expect to
occur in QCD in four dimensions. In the low-energy and strong
coupling limit ($e_{c}>> m_{q}$, $e_{c}$=coupling constant,
$m_{q}$=quark mass ) QCD$_{2}$ has finite-energy soliton solutions
for arbitrary values of $N_{c}$ and $N_{f}$ that can be
interpreted as baryons \cite{frishman}, in close analogy with the
skyrmion interpretation of baryons as solitons in QCD$_{4}$
\cite{witten2}. In this limit the static classical soliton which
describes a baryon in QCD$_{2}$ turns out to be the ordinary
sine-Gordon (SG) soliton. It has been shown  that various aspects
of the low-energy effective QCD$_{2}$ action with unequal quark
masses can be described by the so-called (generalized) sine-Gordon
model (GSG) \cite{jhep5}.

Moreover, it has been proposed that the constituent quark in
QCD$_{4}$ is a topological soliton \cite{kaplan}. These qualitons
(quark solitons), topological excitations with the quantum numbers
of quarks, may provide an accurate description of what is meant by
constituent quarks in QCD. Related to this phenomenon, it has been
found certain static soliton solutions to QCD$_{2}$ that have the
quantum numbers of quarks \cite{ellis111}. They exist only for
quarks heavier than the dimensional gauge coupling ($e_{c} <<
m_{q}$), and have infinite energy, corresponding to the presence
of a string carrying the non-singlet color flux off to spatial
infinity.

On the other hand, the sine-Gordon model (SG) has been studied
over the decades due to its many properties and mathematical
structures such as integrability and soliton solutions. It can be
used as a toy model for non-perturbative quantum field theory
phenomena. In this context, some extensions and modifications of
the SG model deserve attention. An extension taking
multi-frequency
 terms as the potential has been investigated in connection to various physical applications
  \cite{delfino, bajnok, sodano, mussardo}. Another extension defined for multi-fields is the
  so-called generalized
sine-Gordon model (GSG) which has been found in the
 study of the strong/weak coupling sectors of the so-called $sl(N,
\IC)$ affine Toda model coupled to matter fields (ATM) \cite{jmp,
jhep, jhep4}. In connection to these developments, the
bosonization process of the multi-flavor massive Thirring model
(GMT)  provides the quantum version of the (GSG) model
\cite{epjc}. The GSG model
 provides a framework to obtain (multi-)soliton solutions for unequal
mass parameters of the fermions in the GMT sector and study the
spectrum and their interactions. The extension of this picture to
the NC space-time has been addressed (see \cite{jhep12} and
references therein).

It has been conjectured that the low-energy action of QCD$_{2}$
($e>> m_{q}$, $m_{q}$ quark mass and $e$ gauge coupling) might be
related to massive two dimensional integrable models, thus leading
to the exact solution of the strong coupled QCD$_{2}$
\cite{frishman}. In particular, it has been shown that the $sl(2)$
ATM model describes the low-energy spectrum of QCD$_{2}$ (1 flavor
and $N_{c}$ colors) and the exact computation of the string
tension was performed \cite{prd}. A key role has been played by
the equivalence between the Noether and topological currents at
the quantum level. Moreover, one notice that the SU$(n)$ ATM
theory \cite{jmp, jhep} is a $2D$ analogue of the chiral quark
soliton model proposed to describe solitons in QCD$_{4}$
\cite{diakonov}, provided that the pseudo-scalars lie in the
Abelian subalgebra and certain kinetic terms are supplied for
them.

Besides, coupled systems of scalar fields have been investigated
by many authors \cite{Rajaraman, riazi, pogosian, Izquierdo,
Bazeia1, Bazeia2}. One of the motivations was the study of
topological defects in relativistic field theories; since
realistic theories involve more than one scalar field, the
multi-field sine-Gordon
 theories with kink-type exact solutions deserve some attention.
The interest in the study of the classical limit of string theory
on determined backgrounds has recently been greatly stimulated in
connection to integrability. It has been established that the
classical string on $R \times S^2$ is essentially equivalent to
the sine-Gordon integrable system \cite{KP}. More recently, on $R
\times S^3$ background utilizing the Pohlmeyer's reduction it has
been obtained a family of classical string solutions called dyonic
giant magnons which were associated with solitons of complex
sine-Gordon equations \cite{okamura}. String theory on $R\times
S^{N-1}$ is classically equivalent to the so-called $SO(N)$
symmetric space sine-Gordon model (SSG)
 \cite{mikhailov2}.

In the first part of this chapter we study the spectrum of
solitons and kinks  of the GSG model proposed in \cite{jmp, jhep,
epjc} and consider the closely related ATM model from which one
gets the classical GSG model (cGSG) through a gauge fixing
procedure. Some reductions of the GSG model to one-field theory
lead to the usual SG model and to the so-called multi-frequency
sine-Gordon models. In particular, the double (two-frequency)
sine-Gordon model (DSG) appears in a reduction of the $sl(3, \IC)$
GSG model. The DSG theory is a non–integrable quantum field theory
with many physical applications \cite{sodano, mussardo}.

In the ATM model, once a convenient gauge fixing is performed by
setting to constants some spinor bilinears, we are left with two
sectors: the cGSG model which completely decouples from the
spinors and a system of Dirac spinors coupled to the cGSG fields
\cite{jhep4}. In the references \cite{chang, Uchiyama} a
$1+1$-dimensional bag model for quark confinement is considered,
we follow their ideas and generalize for multi-flavor Dirac
spinors coupled to cGSG solitons and kinks. The first reference
considers a model similar to the $sl(2)$ ATM theory, and in the
second one the DSG kink is proposed as an extended hadron model.

In the second part of this chapter we examine the quark soliton
type solutions in QCD$_{2}$. Regarding this phenomenon several
properties of the ATM model deserve careful consideration in view
of the relationships with two-dimensional QCD. For simplicity we
concentrate on the $sl(2)$ ATM model. So, in order to disentangle
the quark solitons one needs to restore the heavy fields, i.e. the
fields associated to the color degrees of freedom. This is done in
two steps. First, by including $N_{c}$ dynamical Dirac spinors
coupled to the Toda field, second by breaking the chiral symmetry
 through certain bilinear terms in the scalar fields of the bosonized
effective Lagrangian. In this way we arrive at a model similar to
the one proposed in \cite{ellis111} in the regime when $m_{q}>>
e_{c}$. We have shown that QCD$_{2}$ has quark soliton solutions
if the quark mass is sufficiently large.

In the next section we define the $sl(3)$ GSG model and study
its properties such as the vacuum structure and the soliton, kink
 and bounce type solutions. In section \ref{atmsec} we consider the
 $sl(3)$ affine Toda model coupled to matter and obtain the cGSG
 model through a gauge fixing procedure. It is discussed the physical soliton
spectrum of the gauge fixed model. In section \ref{topological}
the topological charges are introduced, as well as the idea of
baryons as solitons (or kinks), and the quark confinement
mechanism is discussed. In section \ref{qualitons} we examine the
quark soliton solutions of QCD$_{2}$ and discuss the role played
by the effective $sl(2)$ ATM model. The discussion section
outlines the main results of this contribution  and  some lines of
future research. In appendix \ref{atmapp} we provide the zero
curvature formulation of the $sl(3)$ ATM model.

\section{The GSG model}
\label{model}

The generalized sine-Gordon model (GSG) related to $sl(N)$ is
defined  by \cite{jmp, jhep, epjc} \br \label{GSG} S= \int d^2x
\sum_{i=1}^{N_{f}}\[ \frac{1}{2} (\pa_{\mu} \Phi_{i})^2 + \mu_{i}
\( \mbox{cos} \b_{i} \Phi_{i}-1\)\]. \er The $\Phi_{i}$ fields in
(\ref{GSG})
 satisfy the constraints \br \label{constr0}  \Phi_{p}= \sum_{i=1}^{N-1}
\s_{p\,i} \Phi_{i},\,\,\,\,\,p=N,N+1,...,
N_{f},\,\,\,\,N_{f}=\frac{N(N-1)}{2},\er where
 $\s_{p\,i}$
are some constant parameters and $N_{f}$ is the number of positive
roots of the Lie algebra $sl(N)$. In the context of the Lie
algebraic construction of the GSG system these constraints arise
from the relationship between the positive and simple roots of
$sl(N)$. Thus, in (\ref{GSG}) we have $(N-1)$ independent
fields.

We will consider the $sl(3)$ case with two independent real
fields $\vp_{1, \,2}$, such that \br \label{fields} \Phi_{1}=
2\vp_{1}-\vp_{2};\,\,\,\Phi_{2}= 2\vp_{2}-\vp_{1};\,\,\,\Phi_{3}=
r\, \vp_{1}+ s\, \vp_{2},\,\,\,\,s,r \in \IR \er which must
satisfy the constraint \br \label{constr} \b_{3} \Phi_{3}= \d_{1}
\b_{1} \Phi_{1}+\d_{2} \b_{2} \Phi_{2},\,\,\,\,\b_{i}\equiv
\b_{0}\nu_{i}, \er where $\b_{0},\,\nu_{i},\,\d_{1}, \d_{2}$ are
some real numbers. Therefore, the $sl(3)$ GSG model can be
regarded as three usual sine-Gordon models coupled through the
linear constraint (\ref{constr}).

Taking into account (\ref{fields})-(\ref{constr}) and the fact
that the fields $\vp_{1}$ and $\vp_{2}$ are independent we may get
the relationships \br\label{nus} \nu_{2} \d_{2} = \rho_{0} \nu_{1}
\d_{1} \,\,\,\,\,\nu_{3} =\frac{1}{r+s}(\nu_{1} \d_{1}+\nu_{2}
\d_{2} );\,\,\,\, \rho_{0} \equiv \frac{2s+r}{2r+s} \er

The $sl(3)$ model has a potential density \br V[\vp_{i}] =
\sum_{i=1}^{3} \mu_{i}\(1- \mbox{cos} \b_{i}\Phi_{i}\)
\label{potential}\er

The GSG model has been found in the process of bosonization of the
generalized massive Thirring model (GMT) \cite{epjc}. The GMT
model is a multiflavor extension of the usual massive Thirring
model incorporating massive fermions with current-current
interactions between them. In the $sl(3)$ construction of
\cite{epjc} the parameters $\d_{i}$ depend on the couplings
$\b_{i}$ and  they satisfy certain relationship. This is obtained
by assuming $\mu_{i}
>0$ and the zero of the potential given for $\Phi_{i}=
\frac{2\pi}{\b_{i}} n_{i}$, which substituted into (\ref{constr}) provides \br
\label{deltas} n_{1} \d_{1}+ n_{2} \d_{2}= n_{3},\,\,\,\,n_{i} \in \IZ \er

The last relation combined with (\ref{nus}) gives \br (2r+s)
\frac{n_{1}}{\nu_{1}}+ (2s+r) \, \frac{n_{2}}{\nu_{2}} = 3
\,\frac{n_{3}}{\nu_{3}}\label{nns}.\er

The periodicity of the potential implies an infinitely degenerate
ground state and then the theory supports topologically charged
excitations. A typical potential is plotted in Fig. 1. The vacuum
configuration is related to the fundamental weights (see sections
\ref{atmsec}, \ref{topological} and the Appendix). For the moment,
consider the fields $\Phi_{1}$ and $\Phi_{2}$ and the vacuum
lattice defined by \br (\Phi_{1}\,, \, \Phi_{2})
=\frac{2\pi}{\b_{0}} (\frac{n_{1}}{\nu_{1}}
\,,\,\frac{n_{2}}{\nu_{2}}),\,\,\,\,\,n_{a} \in \IZ.
\label{lattice1}\er

\begin{figure}
\centering \hspace{2.0cm}\scalebox{0.8}{\includegraphics{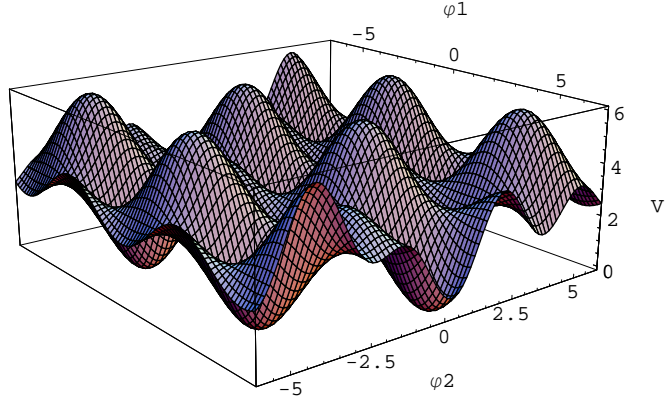}}
\parbox{5in}{\caption{GSG potential $V$ for the parameter
values $\nu_{1}=1/2,\,\,
\delta_{1}=2,\,\,\delta_{2}=1,\,\,\nu_{2}=1,\,\,
r=s=1,\,\,\b_{0}=1,\,\,\mu_{1}=\mu_{2}=1$.}}
\end{figure}

It is convenient to write the equations of motion in terms of the
independent fields $\vp_{1}$ and $\vp_{2}$  \br \pa^2 \vp_{1} &=&
- \mu_{1} \b_{1} \D_{11} \mbox{sin} [\b_{1} ( 2\vp_{1}-\vp_{2})]-
\mu_{2} \b_{2} \D_{12} \mbox{sin}[\b_{2} (2\vp_{2}-\vp_{1})]+
\nonumber\\&& \mu_{3} \b_{3} \D_{13} \mbox{sin} [\b_{3} ( r
\vp_{1} + s \vp_{2})]
\label{eq1}\\
\pa^2 \vp_{2} & = & - \mu_{1} \b_{1} \D_{21} \mbox{sin} [\b_{1} (
2\vp_{1}-\vp_{2})]-\mu_{2} \b_{2} \D_{22} \mbox{sin}[\b_{2}
(2\vp_{2}-\vp_{1})]+ \nonumber\\  && \mu_{3} \b_{3} \D_{23}
\mbox{sin} [\b_{3} ( r \vp_{1}+ s \vp_{2})]\label{eq2}, \er where
\br \nonu A&=&\b_{0}^{2}\nu_{1}^2(4+\d^2+ \d_{1}^2\rh_{1}^2
r^2),\,\,\,\,B=\b_{0}^{2}\nu_{1}^2 (1+4 \d^2 + \d_{1}^2\rh_{1}^2
s^2),\\
\nonumber C&=&\b_{0}^{2}\nu_{1}^2(2+2 \d^2+ \d_{1}^2\rh_{1}^2 r\,
s),\\\nonu
\D_{11}&=&(C-2B)/\D,\,\,\,\,\D_{12}=(B-2C)/\D,\,\,\,\,\D_{13}=(r\,
B+s\, C)/\D,\\\nonu
\D_{21}&=&(A-2C)/\D,\,\,\,\,\D_{22}=(C-2A)/\D,\,\,\,\,\D_{23}=(r\,C+s\,A)/\D\\
 \D&=&C^2-AB,\,\,\,\,\d=\frac{\d_{1}}{\d_{2}} \rh_{0},\,\,\,\,\rh_{1}=\frac{3}{2r+s}\nonu \er

Notice that the eqs. of motion (\ref{eq1})-(\ref{eq2}) exhibit the
symmetry \br\label{symm} \vp_{1} \leftrightarrow
\vp_{2},\,\,\,\,\mu_{1} \leftrightarrow \mu_{2}, \,\,\,\nu_{1}
\leftrightarrow \nu_{2},\,\,\,\d_{1} \leftrightarrow \d_{2},\,\,\,
r \leftrightarrow s \er

Some type of coupled sine-Gordon models have been considered in
connection to various interesting physical problems
\cite{kivshar}. For example a system of two coupled SG models has
been proposed in order to describe the dynamics of soliton
excitations in deoxyribonucleic acid (DNA) double helices
\cite{zhang}. In general these type of equations have been solved
by perturbation methods around decoupled sine-Gordon exact
solitons.

The system of equations (\ref{eq1})-(\ref{eq2}) for  certain
choice of the parameters $r$  and $s$ will be derived in section
\ref{atmsec} in the context of the $sl(3)$ ATM type models, in which
the fields $\vp_{1}$ and $\vp_{2}$ couple to some Dirac spinors in
such a way that the model exhibits a local gauge invariance. The
ATM relevant equations of motion have been solved using a hybrid
of the Hirota and Dressing methods \cite{bueno}. However, in this
reference the physical spectrum of solitons and kinks of the
theory, related to a convenient gauge fixing of the model, have
not been discussed, even though the topological and Noether
currents equivalence has been verified. The appearance of the
so-called tau functions, in order to find soliton solutions in
integrable models, is quite a general result in the both Dressing
and Hirota approaches. In this section,  we will find soliton and
kink type solutions of the GSG model (\ref{eq1})-(\ref{eq2}) and
closely follow the spirit of the above hybrid method approach to
find soliton solutions.

The general tau function for an $n-$soliton solution of the {\sl
gauge unfixed} ATM model has the form \cite{bueno, ferreira} \br \label{tau}
\tau &=& \sum_{p_{1}.....p_{n}=0}^{2}
c_{p_{1}.....p_{n}} \mbox{exp}[p_{1} \G_{i_{1}}(z_{1})+...+p_{n}
\G_{i_{n}}(z_{n})], \\
 &&z_{i}= \g_{i}(x-v_{i}\,
t),\,\,\,\,\,c_{p_{1}.....p_{n}}\in \IC \nonumber \er

Since the GSG model describes the strong coupling sector (soliton
 spectrum) of the ATM model \cite{jmp, jhep} then one can guess the following
Ansatz for the tau functions of the GSG model \br \label{taus}
e^{-i\b_{0}
\frac{\vp_{1}}{2}}=\frac{\tau_{1}}{\tau_{0}},\,\,\,\,\,e^{-i\b_{0}
\frac{\vp_{2}}{2}}=\frac{\tau_{2}}{\tau_{0}}, \er where the tau
functions $\tau_{i}\,(i=0,1,2)$ are assumed to be of the form
(\ref{tau}). We will see that the Ansatz (\ref{taus}) provides
soliton and kink type solutions of the model
(\ref{eq1})-(\ref{eq2}), in this way justifying {\sl a posteriori}
the assumption made for the relevant tau functions.

Assuming that the fields $\vp_{a}\,(a=1,2)$ are real, from
(\ref{taus}) one can write \br  \vp_{1,\,2}&=&\frac{4}{\b_{0}}
\mbox{arctan}[F(\tau_{1,\,2}, \tau_{0})] \label{arct}\\ F&\equiv & \frac{e_{1}
\([Re(\tau_{1,\,2})]^2+[Im(\tau_{1,\,2})]^2\)-\(Re(\tau_{1,\,2})
Re(\tau_{0})+Im(\tau_{1,\,2})
Im(\tau_{0})\)}{\[Im(\tau_{1,\,2})*Re(\tau_{0})-
 Re(\tau_{1,\,2})*Im(\tau_{0})\]},\nonumber \\
e_{1}&=&\pm 1 \label{e1}\er

In terms of the tau functions the system of equations
(\ref{eq1})-(\ref{eq2}) becomes \br \frac{2i}{\b^2_{0}}\[
\frac{\pa^2 \tau_{1}}{\tau_{1}}-\frac{(\pa
\tau_{1})^2}{\tau_{1}^2}-\frac{\pa^2 \tau_{0}}{\tau_{0}}
+\frac{(\pa \tau_{0})^2}{\tau_{0}^2}\]+
 \frac{\b_{1}\mu_{1} \D_{11}}{2i} \[ \frac{(\tau_{2} \tau_{0})^{4\nu_{1}}-\tau_{1}^{8\nu_{1}}}{(\tau_{2} \tau_{0})^{2\nu_{1}} \tau_{1}^{4\nu_{1}}}\]&+&\nonumber \\
 \frac{\b_{2}\mu_{2} \D_{12}}{2i} \[ \frac{(\tau_{1} \tau_{0})^{4\nu_{2}}-\tau_{2}^{8\nu_{2}}}{(\tau_{1} \tau_{0})^{2\nu_{2}} \tau_{2}^{4\nu_{2}}}\]-
 \frac{\b_{3}\mu_{3} \D_{13}}{2i} \[ \frac{(\tau_{0})^{4\nu_{3}(r+s)}-\tau_{1}^{4 r\nu_{3}}\tau_{2}^{4 s\nu_{3}}}{(\tau_{2})^{2s\nu_{3}} (\tau_{1})^{2r\nu_{3}}
 \tau_{0}^{2\nu_{3}(r+s)}}\]&=&0,
 \label{eq1tau}\\
\frac{2i}{\b^2_{0}}\[ \frac{\pa^2 \tau_{2}}{\tau_{2}}-\frac{(\pa \tau_{2})^2}{\tau_{2}^2}-\frac{\pa^2 \tau_{0}}{\tau_{0}}
+\frac{(\pa \tau_{0})^2}{\tau_{0}^2}\]+
 \frac{\b_{1}\mu_{1} \D_{21}}{2i} \[ \frac{(\tau_{2} \tau_{0})^{4\nu_{1}}-\tau_{1}^{8\nu_{1}}}{(\tau_{2} \tau_{0})^{2\nu_{1}} \tau_{1}^{4\nu_{1}}}\]&+&\nonumber \\
 \frac{\b_{2}\mu_{2} \D_{22}}{2i} \[ \frac{(\tau_{1} \tau_{0})^{4\nu_{2}}-\tau_{2}^{8\nu_{2}}}{(\tau_{1} \tau_{0})^{2\nu_{2}} \tau_{2}^{4\nu_{2}}}\]-
 \frac{\b_{3}\mu_{3} \D_{23}}{2i} \[ \frac{(\tau_{0})^{4\nu_{3}(r+s)}-\tau_{1}^{4 r\nu_{3}}\tau_{2}^{4 s\nu_{3}}}{(\tau_{2})^{2s\nu_{3}} (\tau_{1})^{2r\nu_{3}}
 \tau_{0}^{2\nu_{3}(r+s)}}\]&=&0.
 \label{eq2tau}
\er

We will see that the 1-soliton and 1-kink type solutions are
related to half-integer or integer values of the parameters
$\nu_{i}$ and the values $r, s \,=\, 0, 1$. In the next
subsections we write the 1-antisoliton, 1-antikink and bounce type
solutions, and in order to perform the cumbersome computations we
resort to the MAPLE program.

\subsection{One soliton associated to $\vp_{1}$}
\label{s11}

Consider the tau functions
\br \tau_{0}= 1+ i \,d\, \mbox{exp}[\gamma (x-v
t)];\,\,\,\,\tau_{1}=1- i\, d\, \mbox{exp}[\gamma (x-v
t)];\,\,\,\,\tau_{2}= 1+ i\, d\, \mbox{exp}[\gamma (x-v t)]. \nonumber\er

This choice satisfies the system of equations
(\ref{eq1tau})-(\ref{eq2tau}) for the set of parameters \br
\label{para1} \nu_{1}=1/2,\,\,
\delta_{1}=2,\,\,\delta_{2}=1,\,\,\nu_{2}=1,\,\,
\nu_{3}=1,\,\,r=1.\er provided that \br
13\mu_{3}=5\mu_{2}-4\mu_{1},\,\,\,\,
\gamma^2_{1}=\frac{1}{13}(6\mu_{2}+3\mu_{1}).\er

Now, taking  $e_{1}=1$ in Eq. (\ref{e1}) and the relation (\ref{arct}) one has
\br \label{sol1}
\vp_{1}= -\frac{4}{\b_{0}}\mbox{arctan}\{d\,\, \mbox{exp}[\gamma_{1}
(x-v t)]\},\,\,\,\,\vp_{2}=0. \er

This solution is precisely the sine-Gordon 1-antisoliton
associated to the field $\vp_{1}$ with mass $M_{1}=\frac{8
\g_{1}}{\b^2_{0}}$. We plot a soliton of this type in Fig. 3.

\subsection{One soliton associated to $\vp_{2}$}
\label{s22}

Next, let us consider the tau functions \br \nonumber \tau_{0}=1+i\,
d\,\mbox{exp}[\gamma(x-vt)],\,\,\,\,\tau_{1}=
1+i\,d\,\mbox{exp}[\gamma
(x-vt)],\,\,\,\,\tau_{2}=1-i\,d\,\mbox{exp}[\gamma (x-vt)]\er

This set of tau functions solves the system
(\ref{eq1tau})-(\ref{eq2tau}) for the choice of parameters \br
\label{para2} \nu_{1}=1,\,\,
\delta_{1}=1,\,\,\delta_{2}=2,\,\,\nu_{2}=1/2,\,\,
\nu_{3}=1,\,\,s=1\er provided that \br
13\mu_{3}=5\mu_{1}-4\mu_{2},
\,\,\,\,\gamma^2_{2}=\frac{1}{13}(6\mu_{1}+3\mu_{2}) \er

Now, choose $e_{1}=1$ in (\ref{e1}) and through (\ref{arct}) one can get
\br \label{sol2} \vp_{2}= -\frac{4}{\b_{0}}
\mbox{arctan}\{d\,\mbox{exp}[\gamma_{2}(x-vt)]\},\,\,\,\,\vp_{1}=0\er

Similarly, this is the sine-Gordon 1-antisoliton associated to the
field $\vp_{2}$ with mass $M_{2}=\frac{8 \g_{2}}{\b^2_{0}}$ and
its profile is of the type shown in Fig 3.

\subsection{Two one-solitons associated to $\vp\equiv\vp_{1,\,2}$}
\label{s33}

Now, let us consider the tau functions \br \tau_{0} = 1+
i\,d\,\,\mbox{exp}[\gamma(x-v t)],\,\,\,\, \tau_{1} = 1 -
i\,d\,\,\mbox{exp}[\gamma(x-v t)],\,\,\,\,\tau_{2} = 1-
i\,d\,\,\mbox{exp}[\gamma(x-v t)].\nonumber\er

This choice satisfies (\ref{eq1tau})-(\ref{eq2tau}) for
 \br \label{para3} \nu_{1}=1,\, \,\delta_{1}=1/2,\,\,\nu_{2}=1,\,\,
 \delta_{2}=1/2,\,\, \nu_{3}=1/2,\,\,r=s=1,\er  provided that
\br
d^2=1, \,\,\,\,38\gamma^2_{3} = 25\mu_{1}+13\mu_{2}+19\mu_{3}
\er

Now, taking $e_{1}=1$ in (\ref{arct})  one has \br
\vp_{1}&=&\vp_{2}\equiv \hat{\vp}_{1},\,\,\,\,  \label{sol3a}
\\\hat{\vp}_{1} &=& -\frac{4}{\b_{0}} \mbox{arctan}\{d\,\,
\,\mbox{exp}[\gamma_{3}(x-vt)]\}. \label{sol3b}\er

This is a sine-Gordon 1-antisoliton associated to both fields
$\vp_{1,\,2}$ in the particular case when they are equal to each
other. It possesses a mass $M_{3}=\frac{8\g_{3}}{\b_{0}^2}$.

In view of the symmetry (\ref{symm}) we are able to write \br
d^2=1, \,\,\,\,38\gamma^2_{4} = 25\mu_{2}+13\mu_{1}+19\mu_{3}, \er
and then on has another soliton of this type \br
\vp_{1}&=&\vp_{2}\equiv \hat{\vp}_{2},\, \label{sol3c}
\\\hat{\vp}_{2}&=& -\frac{4}{\b_{0}} \mbox{arctan}\{d\,\, \,\mbox{exp}[\gamma_{4}(x-vt)]\}.
\label{sol3d}\er

It possesses a mass $M_{4}=\frac{8\g_{4}}{\b_{0}^2}$. This
1-antisoliton is of the type shown in Fig. 3.

The GSG system (\ref{eq1})-(\ref{eq2}) reduces to the usual SG
equation for each choice of the parameters (\ref{para1}),
(\ref{para2}) and (\ref{para3}), respectively. Then, the
$n-$soliton solutions in each case can be constructed as in the
ordinary sine-Gordon model by taking appropriate tau functions in
(\ref{tau})-(\ref{taus}).

The baryon number associated to each of the above 1-soliton solutions has been computed in connection to QCD$_{2}$, and it takes the same value
$B=N_{c}$ (in this normalization the quark has baryon number $B_{quark}=1$) \cite{jhep5}.

A modified model with rich soliton dynamics is the so-called
stepwise sine-Gordon model in which the system
  parameter depends on the sign of the SG field \cite{riazi1}. It
  would be interesting to consider the above GSG model along the
  lines of this reference.

\subsection{Mass splitting of solitons}
\label{splitt}

It is interesting to write some relations among  the various
soliton masses \br M_{3}^2= \frac{1}{76} (109 M_{2}^2 + 5
M_{1}^2);\,\,\,\, M_{4}^2= \frac{1}{76} (109 M_{1}^2 + 5 M_{2}^2);
\er

If $\mu_{1}=\mu_{2}$  then we have the degeneracy $M_{1}=M_{2}$, and $M_{3}=M_{4}=
\sqrt{3/2} M_{1}$. Notice that if $M_{1}\neq M_{2}$ then $M_{3} < M_{1}+ M_{2}$ and $M_{4} < M_{1}+ M_{2}$,
and the third and fourth solitons are stable in the sense that energy is required to dissociate them.

\subsection{Kinks of the reduced two-frequency sine-Gordon model}
\label{dsg:sec}

In the system (\ref{eq1})-(\ref{eq2}) we perform the following
reduction $\vp \equiv \vp_{1}=\vp_{2}$ such that \br
\Phi_{1}=\Phi_{2},\,\,\,\Phi_{3}= q \, \Phi_{1}, \label{reduc}\er
with $q$ being a real number. Therefore, using the constraint
(\ref{constr}) one can deduce the relationships \br \d_{1} =
\frac{q}{2},\,\,\,\d =1 \label{re1}. \er

Moreover, for consistency of the system of equations
(\ref{eq1})-(\ref{eq2}) we have to impose the relationships \br
\nu_{1}\mu_{1}\D_{11}+ \nu_{2}\mu_{2}\D_{12} &=&
\nu_{1}\mu_{1}\D_{21}+\nu_{2}\mu_{2}\D_{22},\\
\D_{13}&=&\D_{23}.\er

These relations imply \br \label{re2} \d^2 = 1, \,\,\,\mu_{1}=\d
\,\,\mu_{2} \er.

Taking into account the  relations (\ref{re1}) and (\ref{re2})
together with (\ref{nus}) we get \br \mu_{1}=\mu_{2},\,\,\,\,\d =1
,\,\,\,\nu_{1}=\nu_{2},\,\,\,\nu_{3}=\frac{q}{2}
\nu_{1},\,\,r=s=1. \label{param1}\er

Thus the system of Eqs.(\ref{eq1})-(\ref{eq2}) reduce to \br
\label{dsg}\pa^2 \Phi &=& -
\frac{\mu_{1}}{\nu_{1}}\,\mbox{sin}(\nu_{1} \Phi)-\frac{\mu_{3}
\d_{1}}{\nu_{1}} \mbox{sin} (q\, \nu_{1} \Phi),\,\,\,\,\, \Phi
\equiv \b_{0} \vp. \er

This is the so-called {\sl two-frequency sine-Gordon}  model (DSG)
and it has been the subject of much interest in the last decades,
from the mathematical and physical points of view. It encounters
many interesting physical applications, see e.g. \cite{sodano,
mussardo, Uchiyama, kivshar}.

If the parameter $q$ satisfies \br \label{frac1} q =  \frac{n}{m}
\, \in \,  \IQ \er with $m, \,n$ being two relative prime positive
integers, then the potential
$\frac{\mu_{1}}{\nu_{1}^2}(1-\mbox{cos} (\nu_{1} \Phi))
+\frac{\mu_{3}}{2 \nu_{1}^2}(1-\mbox{cos} (q \nu_{1} \Phi))$
associated to the model (\ref{dsg}) is periodic with period \br
\frac{2\pi}{\nu_{1} } m = \frac{2\pi}{q\, \nu_{1} } n. \er

As mentioned above the theory (\ref{dsg}) possesses topological excitations.
The fundamental topological excitations degenerates in the $\mu_{1}=0$
limit to an $n-$soliton state of the relevant sine-Gordon model and similarly
in the limit $\mu_{3}=0$ it will be  an $m$-soliton state. For general values
of the parameters $\mu_{1},\,\mu_{3},\, \d_{1},\,\nu_{1}$ the solitons are in some sense
 ``confined'' inside the topological excitations which become in this form some composite
  objects. On the other hand, if $q \notin \IQ $ then the potential is not periodic,
  so, there are no topologically charged excitations and the solitons are completely
confined \cite{delfino, bajnok}.

The model (\ref{dsg}) in the limit $\mu_{1}=0$  reduces to
\br
\label{eqmu10}
\pa^2 \vp = - \frac{\mu_{3} q}{2 \nu_{1} \b_{0}}\, \mbox{sin} (q \nu_{1} \b_{0} \vp).
\er
For later discussion  we record here the mass of the soliton associated to this equation, \br M_{\mu_{3}} = \frac{8}{(q \nu_{1}\b_{0})^2} \sqrt{q^2 \mu_{3}/2}\label{massmu3}.\er

Correspondingly in the limit  $\mu_{3}=0$ one has \br
\label{eqmu30} \pa^2 \vp = - \frac{mu_{1}}{\nu_{1} \b_{0}}\,
\mbox{sin} (\nu_{1} \b_{0} \vp) \er with associated soliton mass
\br M_{\mu_{1}} = \frac{8}{(\nu_{1}\b_{0})^2} \sqrt{
\mu_{1}}\label{massmu1}\er

Notice that other possibilities to perform  the reduction of type
(\ref{reduc}) encounter some inconsistencies, e.g. the attempt to
implement the reduction $\Phi_{1}=\Phi_{3},\,\,\Phi_{2}= q' \,
\Phi_{1}$ implies $\d_{1,\,2}^2 < 0$ which is a contradiction
since $\d_{1,\,2}$ are real numbers by definition. The same
inconsistency occurs when one tries to reduce the $sl(3)$ GSG
model to a three-frequency SG model. We expect that the three and
higher frequency models \cite{toth} will be related to $sl(N),\,N\ge 4,$ GSG models.

In the following we will provide some kink solutions for
particular set of parameters. Consider \br
\label{paras}\nu_{1}=1/2,\,\,
\d_{1}=\d_{2}=1,\,\,\nu_{2}=1/2,\,\,\,\nu_{3}=1/2 \,\,\,
\mbox{and} \,\,\, q=2,\,n=2,\, m=1\er which satisfy (\ref{param1})
and (\ref{frac1}), respectively. This set of parameters provide
the so-called {\sl double sine-Gordon model} (DSG). Its potential
$-[4 \mu_{1}(\mbox{cos} \frac{\Phi}{2}-1 )+ 2\mu_{3}(\mbox{cos}
\Phi -1)]$ has period $4\pi$ and has extrema at $\Phi = 2\pi
p_{1}$, and\, $\Phi = 4\pi p_{2} \pm 2 \mbox{cos}^{-1}
[1-|\mu_{1}/(2\mu_{3})|]$ with $p_{1},p_{2} \in \IZ$; the second
extrema exists only if $|\mu_{1}/(2\mu_{3})|< 1$. From the
mathematical point of view the DSG model belongs to a
 class of theories with partial integrability \cite{weiss}. Depending on the
 values of the parameters $\b_{0},\, \mu_{1},\,\mu_{3}$ the quantum field theory
 version of the DSG model presents a variety of physical effects, such as the decay
 of the false vacuum, a phase transition, confinement of the kinks and the resonance
 phenomenon due to unstable bound states  of excited kink-antikink states (see \cite{mussardo} and
 references therein). The
semi-classical spectrum of neutral particles in the DSG theory is
investigated in \cite{mussardo1}. Let us mention that the DSG model has recently been in the
center of some controversy regarding the computation of its
semiclassical spectrum, see \cite{mussardo, takacs}.

Interestingly the functions\footnote{These functions are
 obtained by adding the term $\mbox{exp}[2\g (x - vt)]$ to the relevant tau
 functions for one solitons used above. This procedure adds a new method of solving
 DSG which deserve further study.
 The multi-frequency SG equations can be solved through the Jacobi elliptic function
 expansion method, see e.g.  \cite{liu}.}
 \br\nonu \tau_{0}&=&1+i\, d\,
\mbox{exp}[\gamma(x-v t)]+ h\, \mbox{exp}[2\gamma(x-vt)],\,\,\,\\
\tau_{1}&=& 1-i\, d\, \mbox{exp}[\gamma(x-v t)]+ h\,
\mbox{exp}[2\gamma(x-vt)],\er satisfy the equation (\ref{dsg}) for
the parameters (\ref{paras}) provided \br \label{exp11} e^{-i
\Phi/2} &=& \tau_{1}/\tau_{0}
\\ \gamma^2&=&\mu_{1}+2\mu_{3},\,\,\, h=
-\frac{\mu_{1}}{4},\,\,\,\, e_{1}=-1\label{h1}\er

The general solution of this type can be written as \br
\label{gen} \Phi := 4\,
\mbox{arctan}\left[\frac{1}{d}\,\,\frac{1+h\,\,\mbox{exp}[2\gamma(x-vt)]}{\mbox{exp}[\gamma(x-vt)]}\right]
\er

\subsubsection{DSG kink ($h < 0, \,\mu_{i}>0$)}

For the choice of parameters $h < 0, \,\mu_{i}>0 $ in (\ref{h1})
the equation (\ref{gen}) provides \br \label{kink} \vp
:=\frac{4}{\b_{0}} \mbox{arctan}\left[\frac{-2
|h|^{1/2}}{d}\,\,\mbox{sinh}[\gamma_{K}\, (x-vt) +
a_{0}]\right],\,\,\,\,\g_{K}\equiv \pm \sqrt{\mu_{1}+ 2
\mu_{3}},\\ a_{0}=\frac{1}{2} \mbox{ln} |h|.\nonumber\er

This is the DSG 1-kink solution with mass \br M_{K} =
\frac{16}{\b_{0}^2} \g_{K} \left[1 +\frac{\mu_{1}}{\sqrt{2\mu_{3}
(\mu_{1}+ 2\mu_{3})}}\mbox{ln} (\frac{\sqrt{\mu_{1}+ 2\mu_{3}}+
\sqrt{2\mu_{3}}}{\sqrt{\mu_{1}}})\right]. \er

Notice that in the limit $\mu_{1} \rightarrow 0$ the kink mass
becomes $M_{K} = \frac{16}{\b_{0}^2} \sqrt{2\mu_{3}}$, which is
twice the soliton mass (\ref{massmu3}) of the model (\ref{eqmu10})
for the parameters $\nu_{1}=1/2,\, q=2$. Similarly,  in the limit
$\mu_{3} \rightarrow 0$ the kink  mass becomes
$\frac{8}{(\b_{0}/2)^2} \sqrt{\mu_{1}}$, which is the soliton mass
(\ref{massmu1}) of the model (\ref{eqmu30}) for $\nu_{1}=1/2, \,
q=2$; thus in this case the coupling constant is $\b_{0}/2$. As
discussed  above these solitons get in some sense ``confined''
inside the kink if the parameters satisfy $\mu_{i} \neq 0$. The
1-antikink is plotted in Fig. 4. Moreover, the relevant baryon number associated to this DSG kink becomes $B_{kink}=4N_{c}$ \cite{jhep5}.

\subsubsection{Bounce-like solution  ($h>0$, \,$\mu_{1} < 0$)}

For the parameters $h>0$, \,$\mu_{1} < 0$ one gets from
(\ref{gen}) \br \vp :=\frac{4}{\b_{0}}
\mbox{arctan}\left[\frac{2h^{1/2}}{d}\,\,\mbox{cosh}[\gamma
'(x-vt) + a_{0}^{\prime}]\right],\,\,\,\,\,\g '=  2\mu_{3}-
|\mu_{1}|,\,\,\,\,a_{0}^{\prime}=\frac{1}{2} \mbox{ln} h \er

This is the  bounce-like solution and interpolates between the two
vacuum values $2\pi$  and   $4\pi- 2 \mbox{arcos} (1-
|\mu_{1}/2\mu_{3}|)$ and then it comes back. Since $2\pi$ is a
false vacuum position this solution is not related to any stable
particle in the quantum theory \cite{mussardo}. In Fig. 2 we plot
this profile.

\begin{figure} \centering
\hspace{-2.4cm}\scalebox{0.3}{\includegraphics[angle=270]{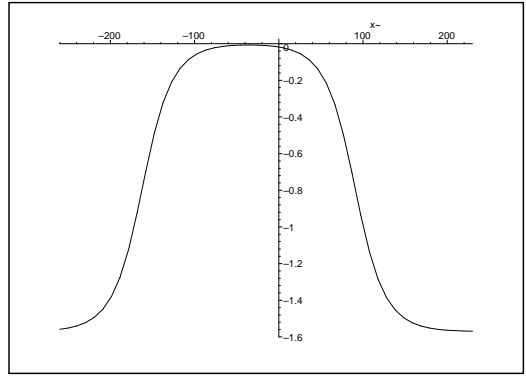}}
\parbox{5in}{\caption{Bounce-like solution ($\frac{\b_{0}}{4} \vp$) plotted for $\mu_{2}=-0.0000001,\,\mu_{3}=0.001,\,d=-2$.}}
\end{figure}\newpage

\section{Classical GSG as a reduced Toda model coupled to matter}
\label{atmsec}

In this section we provide the algebraic construction of the
 $sl(3)$  affine Toda model coupled to matter fields (ATM) and
 closely follows refs. \cite{jhep, bueno, matter}
but the reduction process to arrive at the classical GSG model is
new. The previous treatments of the $sl(3)$ ATM model used
the symplectic and on-shell decoupling methods to unravel the
classical GSG and generalized massive Thirring (GMT) dual theories
describing the strong/weak coupling sectors of the ATM model
\cite{jmp, jhep, annals}. The ATM model describes
 some scalars coupled to spinor (Dirac) fields in which the system of
 equations of motion has a local
gauge symmetry. In this way one includes the spinor sector in the
discussion and conveniently gauge fixing the local symmetry by
setting some spinor bilinears to constants we are able to decouple
the scalar (Toda) fields from the spinors, the final result is a
direct construction of the classical generalized sine-Gordon model
(cGSG) involving only the scalar fields. In the spinor sector we
are left with a system of equations in which the Dirac fields
couple to the cGSG fields.

The zero curvature condition (\ref{zeroc}) gives the following
equations of motion \cite{matter} \br \nonumber \frac{\partial
^{2}\th_{a}}{4i\,e^{\eta}} &=&m^{1}_{\psi}[e^{\eta
-i\phi_{a}}\widetilde{\psi }_{R}^{l}\psi
_{L}^{l}+e^{i\phi_{a}}\widetilde{\psi }_{L}^{l}\psi
_{R}^{l}]+m^{3}_{\psi}[e^{-i\phi_{3}}\widetilde{\psi }
_{R}^{3}\psi _{L}^{3}+e^{\eta +i\phi_{3}}\widetilde{\psi }
_{L}^{3}\psi
_{R}^{3}];\\\,\,\,\,a=1,2 \label{eqnm1} \\
\label{eqnm3} -\frac{\partial ^{2}\widetilde{\nu }}{4}
&=&im^{1}_{\psi}e^{2\eta -\phi_{1}}\widetilde{\psi }_{R}^{1}\psi
_{L}^{1}+im^{2}_{\psi}e^{2\eta -\phi_{2}}\widetilde{\psi
}_{R}^{2}\psi _{L}^{2}+im^{3}_{\psi}e^{\eta
-\phi_{3}}\widetilde{\psi } _{R}^{3}\psi _{L}^{3}+{\bf
m}^{2}e^{3\eta },\,\,
\\
\label{eqnm4} -2\partial _{+}\psi _{L}^{1}&=&m^{1}_{\psi}e^{\eta
+i\phi_{1}}\psi _{R}^{1},\,\,\,\,\,\,\,\,\,\,\,\,\,\,\, -2\partial
_{+}\psi _{L}^{2}\,=\,m^{2}_{\psi}e^{\eta +i\phi_{2}}\psi
_{R}^{2},
\\
\label{eqnm5} 2\partial _{-}\psi _{R}^{1}&=&m^{1}_{\psi}e^{2\eta
-i\phi_{1}}\psi _{L}^{1}+2i
\(\frac{m^{2}_{\psi}m^{3}_{\psi}}{im^{1}_{\psi}}\)^{1/2}e^{\eta
}(-\psi _{R}^{3} \widetilde{\psi }_{L}^{2}e^{i\phi _{2}}-
\widetilde{\psi }_{R}^{2}\psi _{L}^{3}e^{-i\phi_{3}}),
\\
\label{eqnm7} 2\partial _{-}\psi _{R}^{2}&=&m^{2}_{\psi}e^{2\eta
-i\phi_{2}}\psi
_{L}^{2}+2i\(\frac{m^{1}_{\psi}m^{3}_{\psi}}{im^{2}_{\psi}}\)^{1/2}e^{\eta
}(\psi _{R}^{3} \widetilde{\psi }_{L}^{1}e^{i\phi _{1}}+
\widetilde{\psi }_{R}^{1}\psi _{L}^{3}e^{-i\phi _{3}}),
\\
\label{eqnm8} -2\partial _{+}\psi _{L}^{3}&=&m^{3}_{\psi}e^{2\eta
+i\phi _{3}}\psi
_{R}^{3}+2i\(\frac{m^{1}_{\psi}m^{2}_{\psi}}{im^{3}_{\psi}}\)^{1/2}e^{\eta
}(-\psi _{L}^{1}\psi _{R}^{2}e^{i\phi_{2}}+\psi _{L}^{2}\psi
_{R}^{1}e^{i\phi _{1}}),
\\
\label{eqnm9} 2\partial _{-}\psi _{R}^{3}&=&m^{3}_{\psi}e^{\eta
-i\phi_{3}}\psi _{L}^{3},\,\,\,\,\,\,\,\,\,\,\,\, 2\partial
_{-}\widetilde{\psi }_{R}^{1}\,=\,m^{1}_{\psi}e^{\eta
+i\phi_{1}}\widetilde{\psi }_{L}^{1},
\\
\label{eqnm10} -2\partial _{+}\widetilde{\psi }_{L}^{1}
&=&m^{1}_{\psi}e^{2\eta -i\phi_{1}}\widetilde{\psi
}_{R}^{1}+2i\(\frac{m^{2}_{\psi}m^{3}_{\psi}}{im^{1}_{\psi}
}\)^{1/2}e^{\eta }(-\psi _{L}^{2}\widetilde{\psi
}_{R}^{3}e^{-i\phi _{3}}-\widetilde{\psi }_{L}^{3}\psi
_{R}^{2}e^{i\phi_{2}}),
\\
\label{eqnm12} -2\partial _{+}\widetilde{\psi }_{L}^{2}
&=&m^{2}_{\psi}e^{2\eta -i\phi_{2}}\widetilde{\psi
}_{R}^{2}+2i\(\frac{m^{1}_{\psi}m^{3}_{\psi}}{im^{2}_{\psi}}
\)^{1/2}e^{\eta }(\psi _{L}^{1}\widetilde{\psi }_{R}^{3}e^{-i\phi
_{3}}+\widetilde{\psi }_{L}^{3}\psi _{R}^{1}e^{i\phi_{1}}),
\\
\label{eqnm13} 2\partial _{-}\widetilde{\psi
}_{R}^{2}&=&m^{2}_{\psi}e^{\eta+i\phi_{2}}\widetilde{\psi
}_{L}^{2}, \,\,\,\,\,\,\,\,\,\,\,\,\,\,\,\, -2\partial
_{+}\widetilde{\psi }_{L}^{3}\,=\,m^{3}_{\psi}e^{\eta -i\phi
_{3}}\widetilde{\psi }_{R}^{3},
\\
\label{eqnm15} 2\partial _{-}\widetilde{\psi }_{R}^{3}
&=&m^{3}_{\psi}e^{2\eta +i\phi_{3}}\widetilde{\psi
}_{L}^{3}+2i\(\frac{m^{1}_{\psi}m^{2}_{\psi}}{im^{3}_{\psi}}
\)^{1/2}e^{\eta }(\widetilde{\psi} _{R}^{1}\widetilde{\psi
}_{L}^{2}e^{i\phi_{2}}-\widetilde{\psi }_{R}^{2}\widetilde{\psi
}_{L}^{1}e^{i\phi_{1}}),
\\
\label{eqnm16}
\partial^{2}\eta&=&0,
\er where $\phi_{1}\equiv2 \th_{1}-\th_{2},\,\phi_{2}\equiv
2\th_{2}-\th_{1},\,\phi_{3} \equiv \th_{1}+\th_{2}$. Therefore,
one has \br \phi_{3}=\phi_{1}+\phi_{2}\label{phi123}\er

The $\theta$ fields are considered to be in general complex
fields. In order to define the classical generalized sine-Gordon
model we will consider these fields to be real.

Apart from the {\sl conformal invariance} the above equations exhibit the
$\(U(1)_{L}\)^{2}\otimes \(U(1)_{R}\)^{2}$ {\sl left-right local gauge symmetry} \br
\label{leri1}
\th_{a} &\ra& \th_{a} + \xi_{+}^{a}( x_{+}) + \xi_{-}^{a}( x_{-}),\,\,\,\,a=1,2\\
\widetilde{\nu} &\ra& \widetilde{\nu}\; ; \qquad \eta \ra \eta \\
\psi^{i} &\ra & e^{i( 1+ \gamma_5) \Xi_{+}^{i}( x_{+})
+ i( 1- \gamma_5) \Xi_{-}^{i}( x_{-})}\, \psi^{i},\\
\,\,\,\, \widetilde{\psi}^{i} &\ra& e^{-i( 1+ \gamma_5) (\Xi_{+}^{i})( x_{+})-i ( 1-
\gamma_5) (\Xi_{-}^{i})( x_{-})}\,\widetilde{\psi}^{i},\,\,\, i=1,2,3;\label{leri2}
\\
&&\Xi^{1}_{\pm}\equiv \pm \xi_{\pm}^{2} \mp
2\xi_{\pm}^{1},\,\,\Xi^{2}_{\pm}\equiv \pm \xi_{\pm}^{1}\mp
2\xi_{\pm}^{2},\,\,\Xi_{\pm}^{3}\equiv
\Xi_{\pm}^{1}+\Xi_{\pm}^{2}. \nonumber\er

One can get global symmetries for $\xi_{\pm}^{a}=\mp
\xi_{\mp}^{a}=$ constants. For a model defined by a Lagrangian
these would imply the presence of two vector and two chiral
conserved currents. However, it was found only half of such
currents \cite{bueno}. This is a consequence of the lack of a
Lagrangian description for the $sl(3)^{(1)}$ CATM in terms of the
$B$ and $F^{\pm}$ fields (see Appendix). So, the vector current
\br \label{vec} J^{\mu}= \sum_{j=1}^{3} m^{j}_{\psi}
\bar{\psi}^{j}\gamma^{\mu}\psi^{j}\er and the chiral current \br
\label{chi}J^{5\,\mu} = \sum_{j=1}^{3} m^{j}_{\psi}
\bar{\psi}^{j}\gamma^{\mu}\gamma_{5} \psi^{j}+ 2 \partial_{\mu}
(m^{1}_{\psi}\th_{1}+m^{2}_{\psi}\th_{2})\er are conserved \br
\label{conservation} \pa_{\mu} J^{\mu}=0,\,\,\,\,\,\pa_{\mu}
J^{5\, \mu}=0\er

The conformal symmetry is gauge fixed by setting \br \eta =
\mbox{const}. \label{eta}\er

The off-critical model obtained in this way exhibits the vector
and topological currents equivalence \cite{matter, annals} \br
\label{equivalence} \sum_{j=1}^{3} m^{j}_{\psi}
\bar{\psi}^{j}\gamma^{\mu}\psi^{j} \equiv \epsilon^{\mu
\nu}\partial_{\nu}
(m^{1}_{\psi}\theta_{1}+m^{2}_{\psi}\theta_{2}),\,\,\,\,\,\,\,
m^{3}_{\psi}=m^{1}_{\psi}+ m^{2}_{\psi},\,\,\,\,m^{i}_{\psi}>0.
\er

Moreover, it has been shown that the soliton type solutions are in
the orbit of the vacuum $\eta=0$.

In the next steps we implement the reduction process to get the
cGSG model through a gauge fixing of the ATM theory. The local
symmetries (\ref{leri1})-(\ref{leri2}) can be gauge fixed through
\br \label{gf} i \bar{\psi}^{j}\psi^{j}= i
A_{j}=\mbox{const.};\,\,\,\,\,\,\bar{\psi}^{j}\gamma_{5}\psi^{j}
=0. \er

From the gauge fixing (\ref{gf}) one can write the following
bilinears
 \br
 \label{bilinears}
 \widetilde{\psi}_{R}^{j} \psi_{L}^{j} +
 \widetilde{\psi}_{L}^{j}\psi_{R}^{j}=0,\,\,\,\,\,j=1,2,3;
 \er
so,  the eqs. (\ref{gf})  effectively comprises three gauge fixing
 conditions.

It can be directly verified that the gauge fixing (\ref{gf})
preserves the currents conservation laws (\ref{conservation}),
i.e. from the equations of motion (\ref{eqnm1})-(\ref{eqnm16})
 and the gauge fixing (\ref{gf}) together with (\ref{eta}) it is possible
 to obtain the currents conservation laws (\ref{conservation}).

Taking into account the constraints (\ref{gf}) in the scalar
sector, eqs. (\ref{eqnm1}), we arrive at
 the following system
of equations (set $\eta=0$)\br \label{sys1} \pa^2 \th_{1} &=&
M^{1}_{\psi}\,
\mbox{sin} \phi_{1} + M_{\psi}^{3}\, \mbox{sin} \phi_{3},\\
\label{sys2}\,\,\,\pa^2 \th_{2} &=& M^{2}_{\psi}\, \mbox{sin}
\phi_{2} + M^{3}_{\psi}\, \mbox{sin} \phi_{3},\,\,\,\,
M^{i}_{\psi} \equiv 4 A_{i}\, m_{\psi}^{i},\,\,\,\,i=1,2,3.\er

Define the fields $\vp_{1},\,\vp_{2}$ as \br
\label{transf1}\vp_{1} &\equiv& a \th_{1} +
b\th_{2},\,\,\,\,\,\,\,\,
a=\frac{4\nu_{2}-\nu_{1}}{3\b_{0}\nu_{1}\nu_{2}},\,\,\,d=\frac{4\nu_{1}-\nu_{2}}{3\b_{0}\nu_{1}\nu_{2}}
\\\vp_{2}&\equiv& c \th_{1} + d
\th_{2},\,\,\,\,\,\,\,\, b=-c=\frac{2(\n_{1}-\nu_{2})}{3\b_{0}
\nu_{1}\nu_{2}}, \,\,\,\, \nu_{1}, \nu_{2} \in \IR
\label{transf2}\er

Then, the system of equations (\ref{sys1})-(\ref{sys2}) written in
terms of the fields $\vp_{1,\,2}$ becomes \br  \pa^2 \vp_{1} &=& a
M^{1}_{\psi}\, \mbox{sin} [\b_{0}\nu_{1}(2\vp_{1}-\vp_{2})] + b
M^{2}_{\psi}\, \mbox{sin} [\b_{0}\nu_{2}(2\vp_{2}-\vp_{1})] +
\nonumber \\&&(a+b) M^{3}_{\psi}\, \mbox{sin}
\b_{0}[(2\nu_{1}-\nu_{2})\vp_{1}+(2\nu_{2}- \nu_{1})\vp_{2})],
\label{cgsg1} \\
\nonumber \pa^2 \vp_{2} &=& c M^{1}_{\psi}\, \mbox{sin}
[\b_{0}\nu_{1}(2\vp_{1}-\vp_{2})] + d M^{2}_{\psi}\, \mbox{sin}
[\b_{0}\nu_{2}(2\vp_{2}-\vp_{1})] +
\\&& (c+d) M^{3}_{\psi}\, \mbox{sin} \b_{0}[(2\nu_{1}-\nu_{2})\vp_{1}+(2\nu_{2}-
\nu_{1})\vp_{2})]\label{cgsg2} \er

The system of equations above considered for real fields
$\vp_{1,\,2}$ as well as for real parameters $M_{\psi}^{i}, a, b,
c, d, \b_{0}$ defines the {\sl classical generalized sine-Gordon
model} (cGSG).
 Notice that this classical
version of the GSG model derived from the ATM theory is a submodel
of the GSG model (\ref{eq1})-(\ref{eq2}), defined in section
\ref{model}, for the particular parameter values
 $r=\frac{2\nu_{1}-\nu_{2}}{\nu_{3}},\,
 s=\frac{2\nu_{2}-\nu_{1}}{\nu_{3}}$ and the convenient identifications of the
 parameters in the coefficients of the sine functions of the both  models.

The following reduced models can be obtained from the system
(\ref{cgsg1})-(\ref{cgsg2}):

i){\sl SG submodels}\\
i.1) For $\nu_{2}=2\nu_{1}$ \,one has  $M^{1}_{\psi} =
M^{2}_{\psi}$ and the system \\ $\vp_{2}=0$,\,\, $\pa^{2}\vp_{1}=
M^{1}_{\psi} \frac{3\nu_{1}}{\b_{0}}\, \mbox{sin}\,\b_{0} 2\nu_{1}
\vp_{1}$.
\\i.2) For $\nu_{1}=2\nu_{2}$ \,one has  $ M^{1}_{\psi}= M^{2}_{\psi}$
and the system \\ $\vp_{1}=0$,\,\, $\pa^{2}\vp_{2}= M^{2}_{\psi}
\frac{3\nu_{2}}{\b_{0}}\, \mbox{sin}\, \b_{0} 2\nu_{2} \vp_{2}$.
\\ i.3) For $\nu_{2}=\nu_{1}\equiv \nu$\, and\,$\vp_{1}=\vp_{2}\equiv \hat{\vp}_{A},\, (A=1,2) $, one gets the
sub-models

i.3a) $M^{1}_{\psi}=M^{2}_{\psi},\,M^{3}_{\psi}=0$,\,\, $\pa^{2}
\hat{\vp}_{1}= a M^{1}_{\psi} \, \mbox{sin}\, \b_{0} \nu
\hat{\vp}_{1},$

 i.3b) $M^{1}_{\psi}=M^{2}_{\psi}=0$,\,\, $\pa^{2} \hat{\vp}_{2}=
a M^{3}_{\psi} \, \mbox{sin}\, \b_{0} \nu \hat{\vp}_{2}.$

ii) {\sl DSG sub-model}\\ For $\nu_{1}=\nu_{2}$ \,and \, $
M^{1}_{\psi}= M^{2}_{\psi}$ one gets the sub-model
$\vp_{1}=\vp_{2}\equiv \vp$,\,\, $\pa^{2}\vp= a M^{1}_{\psi} \,
\mbox{sin}\, \b_{0} \nu_{1} \vp + a M^{3}_{\psi}\, \mbox{sin}\,
2\b_{0} \nu_{1} \vp$.

The sub-models i.1)-i.2) each one contains the ordinary
sine-Gordon model (SG) and they were considered in the subsections
\ref{s11} and \ref{s22}, respectively; the sub-model i.3) supports
two SG models
 with different soliton masses which must correspond to the construction in subsection \ref{s33};
 and the ii) case defines the double
sine-Gordon model (DSG) studied in subsection \ref{dsg:sec}. Other
meaningful reductions are possible arriving at either SG or DSG
model. Notice that the reductions above are particular cases of
the sub-models in subsections \ref{s11}, \ref{s22}, \ref{s33} and
 \ref{dsg:sec}, respectively, for relevant parameter identifications.

The spinor sector in view of the gauge fixing (\ref{gf}) can be
parameterized conveniently as \br \left(
\begin{array}{c}
 \psi_{R}^{j}\\\psi_{L}^{j}
\end{array}\right)= \left(
\begin{array}{c}
\sqrt{A_{j}/2}\,u_{j} \\i \sqrt{A_{j}/2}\,\frac{1}{v_{j}}
\end{array}\right);\,\,\,\,\left(
\begin{array}{c}
 \widetilde{\psi}_{R}^{j}\\\widetilde{\psi}_{L}^{j}
\end{array}\right)= \left(
\begin{array}{c}
\sqrt{A_{j}/2}\, v_{j} \\-i \sqrt{A_{j}/2}\, \frac{1}{u_{j}}
\end{array}\right).\label{paramet}\er

Therefore, in order to find the spinor field solutions one can
solve the eqs. (\ref{eqnm4})-(\ref{eqnm15}) for the fields $u_{j},
v_{j}$ for each solution given for the cGSG fields $\vp_{1,\,2}$
of the system (\ref{cgsg1})-(\ref{cgsg2}).

\subsection{Physical solitons and kinks of the ATM model}

The main feature of the one `solitons' constructed in \cite{bueno}
is that for each positive root of $sl(3)$ there corresponds one
soliton species associated to the fields
$\phi_{1},\,\phi_{2},\,\phi_{3}$, respectively. The relevant
solutions for the spinor fields together with the 1-`solitons'
satisfy the relationship (\ref{equivalence}). The class of
$2$-`soliton' solutions of $sl(3)$ ATM obtained in \cite{bueno}
behave as follows:\, i) they are given by 6 species associated to
the pair $(\a_{i},\a_{j}),\, i\le j;\,\, i,j=1,2,3$; where the
$\a$'s are the positive roots of $sl(3)$ Lie algebra. Each species
$(\a_{i},\a_{i})$ solves the $sl(2)$ ATM submodel\footnote{$sl(2)$
ATM gauge unfixed $2-$'solitons' satisfy an analogous eq. to
\rf{equivalence}. Moreover, for $\vp$ real and
$\widetilde{\psi}=\pm (\psi)^{*}$ one has, soliton-soliton $SS$,
$SS$ bounds and no $S\bar{S}$ ($S=$soliton,\,
$\bar{S}$=anti-soliton) bounds \cite{nucl1} associated to the
field $\vp$.}. ii) they satisfy the $U(1)$ vector and topological
currents equivalence \rf{equivalence}. However, the possible kink
type solutions associated in a {\sl non-local} way to the spinor
bilinears and the relevant gauge fixing of the local symmetry
(\ref{leri1})-(\ref{leri2}) have not been discussed in the
literature. In order to consider the physical spectrum of solitons
and study its properties, such as their masses and scattering time
delays, it is mandatory to take into account these questions which
are related to the counting of the true physical degrees of
freedom of the theory. Therefore, one must consider the possible
soliton type solutions  associated to each spinor bilinear. The
relation between  this type of `solitons', say $\hat{\phi}_{j}$,
and their relevant fermion bilinears must be non-local as
suggested by the equivalence equation (\ref{equivalence}). So, we
may have soliton solutions of type \br \hat{\phi}_{j} = \int^{x}\,
dx^{\prime} \, \bar{\psi}^{j}\g^{0}\psi^{j} ,\,\,\,\,\,\,j=1,2,3
\label{nonlocal}\er

At this stage one is able to enumerate the physical 1-soliton
(1-antisoliton) spectrum associated to the gauge fixed ATM model.
In fact, we have three 'kinks' and their corresponding
'anti-kinks' associated to the fields $\phi_{i}$ (i=1,2,3), and
three kink and antikink pairs of type $\hat{\phi}_{j},\,j=1,2,3$.
Thus, we have six kink and their relevant antikink solutions, but
in order to record the physical soliton and anti-soliton
excitations one must take into account the four constraints
(\ref{phi123}) and (\ref{bilinears}). Therefore, we expect to find
four pairs of soliton and anti-soliton physical excitations in the
spectrum. This feature is nicely reproduced in the cGSG sector of
the ATM model; in fact, in the last section we were able to write
four usual sine-Gordon models as possible reductions of the cGSG
model. Namely, one soliton associated to the fields $\vp_{1},\,
\vp_{2}$, respectively (subsections \ref{s11} and \ref{s22}) and
1-solitons associated to the field $\vp_{1}=\vp_{2}\equiv
\vp_{A},\,A=1,2$, respectively (subsection \ref{s33}). In the
2-kink (2-antikink) sector a similar argument will provide us ten
physical 2-solitons and their relevant 2-antisoliton excitations,
i.e. six pairs of 2-kink and 2-antikink solutions of type $\phi$
 and $\hat{\phi}$, respectively, which give twenty four excitations, and
 taking into account  the
 constraints (\ref{phi123}) and (\ref{bilinears}) we are left with
 ten pairs of 2-solitons and 2-antisolitons. In fact, these
 ten 2-solitons correspond to the pairs we can form with the four
 species of 1-solitons in all possible
 ways. The same argument holds for the corresponding ten
 2-antisolitons.

 In this way the system (\ref{cgsg1})-(\ref{cgsg2}) gives rise to a
 richer (anti)soliton spectrum and dynamics than the $\theta_{a}$ field
'soliton' type solutions  of the gauge unfixed model
(\ref{eqnm1})-(\ref{eqnm15}) found in \cite{bueno}. Regarding this
issue let us notice that in the procedure followed in ref.
\cite{bueno} the local symmetry (\ref{leri1})-(\ref{leri2}) and
the relevant gauge fixing has not been considered explicitly,
therefore their 'solitons' do not correspond to the GSG solitons
obtained above.

Notice that the tau functions in section \ref{model} possess the
function $\g (x-v t)$ in their exponents, whereas the
corresponding ones in the ATM theory  have two times this function
\cite{bueno, nucl1}. This fact is reflected in the GSG soliton
solutions which are two times the relevant solutions of the ATM
model.  It has been observed already in the $sl(2)$ case that the
 $\theta$ `soliton' of the gauge unfixed $sl(2)$ ATM model (see eq. (2.22) of
\cite{nucl1}) is half the soliton of the usual SG model.

\section{Topological charges, baryons as solitons and confinement}
 \label{topological}

In this section we will examine the vacuum configuration of the
cGSG model and the equivalence between the $U(1)$ spinor current
and  the topological current (\ref{equivalence}) in the gauge
fixed model and verify that the charge associated to the $U(1)$
current gets confined inside the solitons and kinks of the GSG
model obtained in section \ref{model}.

It is well known that in 1 + 1 dimensions the topological current
is defined as $J_{\mbox{top}}^{\mu} \sim
\epsilon^{\mu\nu}\pa_{\nu}\Phi$, where $\Phi$ is some scalar
field. Therefore, the topological charge is $ Q_{\mbox{top}} =
\int J_{\mbox{top}}^{0} dx \sim  \Phi(+ \infty) - \Phi(-\infty) $.
In order to introduce a topological current we follow the
construction adopted in Abelian  affine Toda models, so we define
the field \br \theta = \sum_{a=1}^{2} \frac{2 \a_{a}}{\a^2_{a}}
\theta_{a} \er where $\a_{a},\,a=1,2$, are the simple roots of
$sl(3)$. We then have that $\theta_{a} = (\theta | \l_{a})$,
where $\l_{a}$ are the fundamental weights of $sl(3)$ defined
by the relation \cite{humphreys} \br 2 \frac{( \a_{a} |
\l_{b})}{(\a_{a}|\a_{a})}= \d_{ab}. \er

The fields $\phi_{j}$ in the equations
(\ref{eqnm1})-(\ref{eqnm15}) written as the combinations $(\theta|
\a_{j}), \, j=1,2,3$, where the $\a_{j}'s$ are the positive roots
of  $sl(3)$, are invariant under the transformation \br
\theta \rightarrow \theta + 2\pi
\mu\,\,\,\,\,\,\,\,&\mbox{or}&\,\,\,\,\,\,\,\,\phi_{j}\rightarrow
\phi_{j} + 2\pi
(\mu|\a_{j}),\label{transdiscret}\\
\mu &\equiv& \sum_{n_{a}\in \IZ} n_{a} \frac{2
\vec{\l}_{a}}{(\a_{a}|\a_{a})},\label{weight}\er where $\mu$ is a
weight vector of $sl(3)$, these vectors satisfy $(\mu
|\a_{j}) \in \IZ $ and form an infinite discrete lattice called
the weight lattice \cite{humphreys}. However, this weight lattice
does not constitute the vacuum configurations of the ATM model ,
since in the model described by (\ref{eqnm1})-(\ref{eqnm16}) for
any constants $\theta_{a}^{(0)}$ and $\eta^{(0)}$ \br
\label{vacuum} \psi_{j} = \widetilde{\psi}_{j}
=0,\,\,\theta_{a}=\theta_{a}^{(0)},\,\,\eta=\eta^{(0)},\,\,\,\widetilde{\nu}
= - {\bf{m}}^{2}e^{\eta^{(0)}} x^{+} x^{-}\er is a vacuum
configuration.

We will see that the topological charges of the physical
one-soliton solutions of (\ref{eqnm1})-(\ref{eqnm16}) which are
associated to the new fields $\vp_{a},\,a=1,2,$ of the cGSG model
(\ref{cgsg1})-(\ref{cgsg2}) lie on a modified lattice which is
related to the weight lattice by re-scaling the weight vectors. In
fact, the eqs. of motion (\ref{cgsg1})-(\ref{cgsg2}) for the field
defined by $\vp \equiv \sum_{a=1}^{2} \frac{2 \a_{a}}{\a^2_{a}}
\vp_{a},\,$ such that $\vp_{a} = (\vp | \l_{a})$, are invariant
under the transformation \br \vp \rightarrow \vp +
\frac{2\pi}{\b_{0}}\sum_{a=1}^{2}
 \frac{q_{a}}{\nu_{a}} \frac{2
\l_{a}}{(\a_{a}|\a_{a})},\,\,\,q_{a} \in \IZ.\er

So, the vacuum configuration is formed by an infinite discrete
lattice related to the usual weight lattice by the relevant
re-scaling of the fundamental weights $\l_{a}\rightarrow
\frac{1}{\nu_{a}} \l_{a}$. The vacuum lattice can be given by the
points in the plane $\vp_{1}$\, x\, $\vp_{2}$ \br (\vp_{1}
\,,\,\vp_{2}) = \frac{2\pi}{3\b_{0}}(\frac{2 q_{1}}{\nu_{1}} +
\frac{ q_{2}}{\nu_{2}}\,,\,\frac{ q_{1}}{\nu_{1}}+\frac{2
q_{2}}{\nu_{2}}),\,\,\,\,\,q_{a} \in \IZ.\er

In fact, this lattice is related to one in eq. (\ref{lattice1})
through appropriate parameter identifications. We shall define the
topological current and charge, respectively, as \br
 J_{\mbox{top}}^{\mu} = \frac{\b_{0}}{2\pi} \epsilon^{\mu\nu} \pa_{\nu}
 \vp,\,\,\,\,\,\,\,\,
 Q_{\mbox{top}} = \int dx J_{\mbox{top}}^{0} = \frac{\b_{0}}{2\pi}
 [\vp (+ \infty) -\vp(\infty)]. \label{topcurrcgsg}\er

Taking into account the cGSG fields (\ref{cgsg1})-(\ref{cgsg2})
and the spinor parameterizations (\ref{paramet}) the currents
equivalence (\ref{equivalence}) of the  ATM model takes the form
 \br
\label{equivalence1} \sum_{j=1}^{3} m^{j}_{\psi}
\bar{\psi}^{j}\gamma^{\mu}\psi^{j} \equiv \epsilon^{\mu
\nu}\partial_{\nu} ({\zeta}^{1}_{\psi}\,
\vp_{1}+\zeta^{2}_{\psi}\, \vp_{2}), \er where $\zeta^{1}_{\psi}
\equiv \b_{0}^{2} \nu_{1}\nu_{2} (m^{1}_{\psi} d + m^{2}_{\psi}
b),\,\,\, \zeta^{2}_{\psi} \equiv \b_{0}^{2}\nu_{1}\nu_{2}
(m^{2}_{\psi} a -  m^{1}_{\psi} b)$ and the spinors are understood
to be written in terms of the fields $u_{j}$\, and \,$v_{j}$ of
(\ref{paramet}).

Notice that the topological current in (\ref{equivalence1}) is the
projection of (\ref{topcurrcgsg}) onto the vector
$\frac{2\pi}{\b_{0}} \( \zeta_{\psi}^{1} \l_{1} + \zeta_{\psi}^{2}
\l_{2}\)$.

As mentioned in section \ref{atmsec} the gauge fixing (\ref{gf})
preserves the currents conservation laws (\ref{conservation}).
Moreover, the cGSG model was defined for the off critical ATM
model obtained after setting $\eta=\mbox{const}.=0$. So, for the
gauge fixed model it is expected to hold the currents equivalence
relation (\ref{equivalence}) written for the spinor
parameterizations  $u_{j}, v_{j}$ and the fields $\vp_{1,2}$ as is
presented in eq. (\ref{equivalence1}). Therefore, in order to
verify the $U(1)$ current confinement it is not necessary to find
the explicit solutions for the spinor fields. In fact, one has
that the current components are given by relevant partial
derivatives of the linear combinations of the field solutions,
$\vp_{1, 2}$, i.e. \, $J^{0}=\sum_{j=1}^{3} m^{j}_{\psi}
\bar{\psi}^{j}\gamma^{0}\psi^{j} \, = \, \partial_{x}
({\zeta}^{1}_{\psi}\, \vp_{1}+\zeta^{2}_{\psi}\,
\vp_{2})$\,and\,$J^{1}=\sum_{j=1}^{3} m^{j}_{\psi}
\bar{\psi}^{j}\gamma^{1}\psi^{j} \, = \, -\partial_{t}
({\zeta}^{1}_{\psi}\, \vp_{1}+\zeta^{2}_{\psi}\, \vp_{2})$. In
particular the current components $J^{0}, J^{1}$ and their
associated scalar field solutions are depicted in Figs. 3 and 4,
respectively, for antisoliton and antikink solutions.

It is clear that the charge density related to this $U(1)$ current
can only take significant values on those regions where the
$x-$derivative of the fields $\vp_{1,2}$ are non-vanishing. That
is one expects to happen with the bag model like confinement
mechanism in quantum chromodynamics (QCD). As we have seen the
soliton and kink solutions of the GSG theory are localized in
space, in the sense that the scalar fields interpolate between the
relevant vacua in a limited region of space with a size determined
by the soliton masses. The spinor $U(1)$ current gets the
contributions from all the three spinor flavors. Moreover, from
the equations of motion (\ref{eqnm4})-(\ref{eqnm15})
 one can obtain nontrivial spinor solutions different from vacuum
(\ref{vacuum}) for each set of scalar field solutions $\vp_{1},
\vp_{2}$. For example, the solution $\vp_{1}=$soliton, $\vp_{2}=0$
in section \ref{s11} implies $\phi_{1}=\vp_{1},\,
\phi_{2}=-\vp_{1},\,\phi_{3}=0$ which substituting into the spinor
equations of motion (\ref{eqnm4})-(\ref{eqnm15}) will give
nontrivial spinor field solutions. Therefore, the ATM model of
section \ref{atmsec} can be considered as a multiflavor
generalization of the two-dimensional hadron model proposed in
\cite{chang, Uchiyama}. In the last reference a scalar field is
coupled to a spinor such that the
 DSG kink arises as a model for hadron and  the quark field
 is confined inside the bag.

\begin{figure}
\centering
\hspace{-2.4cm}\scalebox{0.4}{\includegraphics[angle=0]{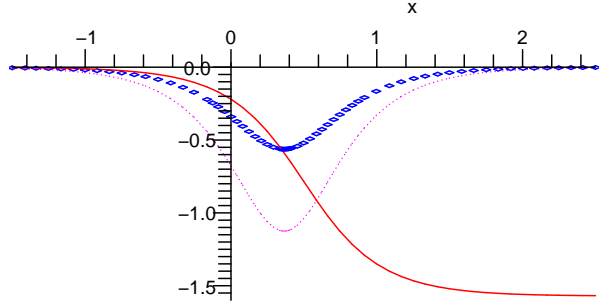}}
\parbox{5in}{\caption{1-antisoliton and
confined current $J^{\mu}$. The solid curve is the 1-antisoliton
($\frac{\b_{0}}{4} \vp$), the dashdotted curve is $J^{0}$ and the
curve with losangles is $J^{1}$. For $t=1,\,\mu_{1}=\mu_{2}=1,
d=1.5, v=0.05, \b_{0}=0.5,\,
m_{\psi}^{1}=m_{\psi}^{2}=1$,\,$\nu_{1}=1,\,\d_{1}=1,\,\d_{2}=2$.}}
\end{figure}

\begin{figure}
\centering \hspace{-2.4cm}\scalebox{0.4}{\includegraphics{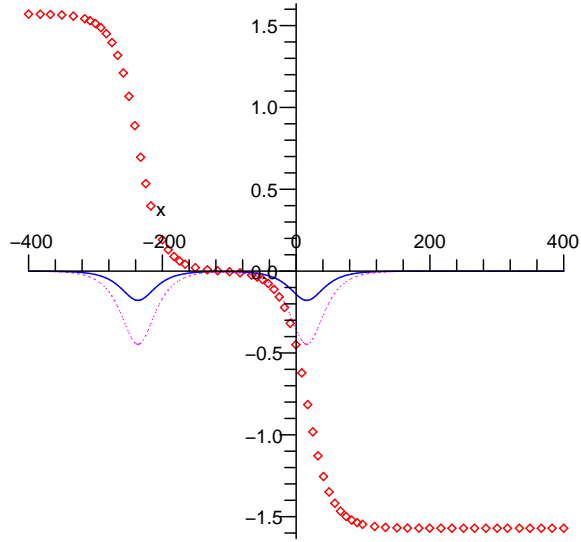}}
\parbox{5in}{\caption{DSG kink solution and confined current $J^{\mu}$. The curve with
losangles is the antikink ($\frac{\b_{0}}{4} \vp$), the dashdotted
line is $J^{0}$, the solid curve is $J^{1}$. For $t=1,\,\b_{0}=
10^8,\, m^{1,
2}_{\psi}=\mu_{1}=-0.0000001,\,\,\mu_{3}:=0.001,\,\,d=2,\,\d_{1}=\d_{2}=1,\,\nu_{1}=1/2
$.}}
\end{figure}\newpage

\section{Qualitons or quark solitons in two-dimensional QCD}
\label{qualitons}

Several properties of the ATM model deserve careful consideration
in view of the relationships with two-dimensional QCD. In
particular, it has been shown that the $sl(2)$ ATM model describes
the low-energy spectrum of QCD$_{2}$ ($1$ flavor and $N_{c}$
colors) \cite{prd}. In the context of bosonized QCD$_{2}$ the
appearance of soliton solutions that have
the quantum numbers of quarks as constituent of hadrons has  been considered \cite{ellis111}. So, one can inquire
about these type of quark solitons in the context of the ATM model
description of QCD$_{2}$. Since the ATM model describes the
low-energy effective action in the strong coupling limit of
QCQ$_{2}$, in order to disentangle the quark solitons  one needs
to restore, in some way, the heavy fields, i.e. the fields
associated to the color degrees of freedom. For simplicity we
choose the $sl(2)$ case in the following developments.

The Lagrangian of the $sl(2)$ ATM model is defined by \cite{annals, nucl1, nucl}
\begin{eqnarray}
\label{atm}
 \frac{1}{k}{\cal L} = \frac{1}{4} \partial_{\mu} \varphi \, \partial^{\mu} \varphi
+ i  {\bar{\psi}} \gamma^{\mu} \partial_{\mu} \psi
- m_{\psi}\,  {\bar{\psi}} \,
e^{2i\varphi\,\gamma_5}\, \psi,
\end{eqnarray}
where $k=\frac{\kappa}{2\pi}$,\, ($\kappa \in \hbox{\sf Z}$), $\varphi$ is a real field,
 $m_{\psi}$ is a mass parameter, and $\psi$ is a Dirac spinor. Notice that ${\bar{\psi}}
  \equiv {\widetilde{\psi}}^{T} \,\gamma_0$. We shall take ${\widetilde{\psi}}=e_{\psi}
   \psi^{*}$ \cite{nucl1}, where $e_{\psi}$ is a real dimensionless constant.
The conformal version (CATM) of (\ref{atm}) has been constructed in \cite{matter}.
The integrability properties and the reduction processes: WZNW$\rightarrow$
 CATM $\rightarrow$ ATM $\rightarrow$ sine-Gordon(SG) $+$ free field, have been
 considered \cite{annals, nucl1, nucl} . The $sl(n)$ ATM exhibits a generalized
 sine-Gordon/massive Thirring correspondence \cite{jmp}. Moreover, (\ref{atm})
 exhibits mass generation despite chiral symmetry \cite{witten} and confinement
 of fermions in a self-generated potential \cite{nucl1, chang}.

The Lagrangian is invariant under $\varphi \rightarrow \varphi + n \pi$, thus the
topological charge, $Q_{\mbox{topol.}} \equiv \int \, dx \, j^0 ,
\,j^{\mu} =  \frac{1}{\pi}\epsilon^{\mu\nu} \partial_{\nu} \, \varphi$, can assume
nontrivial values. A reduction is performed imposing the constraint
\begin{eqnarray}
\label{equiv}
\frac{1}{2\pi}\epsilon^{\mu\nu} \partial_{\nu} \, \varphi=
\frac{1}{\pi} \bar \psi \gamma^\mu  \psi,
\end{eqnarray}
where $J_{\mu}= \bar \psi \gamma^\mu  \psi$ is the $U(1)$ Noether current. In fact,
the soliton type solutions satisfy this relationship \cite{nucl1}.

The Eq. (\ref{equiv}) implies $\psi^{\dagger} \psi \sim \partial_{x} \varphi$,  thus
the Dirac field is confined to live in regions where the field  $\varphi$ is not constant.
The $1(2)-$soliton(s) solution(s) for $\varphi$ and $\psi$ are of the sine-Gordon (SG) and
massive Thirring (MT) types, respectively; they satisfy (\ref{equiv}) for $|e_{\psi}| = 1$,
and so are solutions of the reduced model \cite{nucl1}. Similar results hold in $sl(n)$
ATM \cite{bueno, jmp}.

The equivalence (\ref{equiv}) for multisolitons describes,
 $\varphi = \varphi_{N}$ [$Q_{\mbox{topol}}= N\, \mbox{sign}(e_{\psi})$] and
 $\Psi$ $N-$solitons of the SG and MT type, respectively. Asymptotically one can write
\begin{eqnarray}
\label{equiv2}
\frac{1}{2\pi}\epsilon^{\mu\nu} \partial_{\nu} \, \varphi_{N} \approx
\sum_{a=1}^{N} \frac{1}{\pi} \bar \psi_{a} \gamma^\mu  \psi_{a},
\end{eqnarray}
where the $\psi_{a}$'s are the solutions for the individual localized lowest energy
fermion states. In fact, (\ref{equiv2}) encodes the classical SG/MT correspondence
\cite{orfanidis}. Thus, the ATM model can accomodate $N_{c}=N-$fermion confined states
with internal `color' index $a$ \cite{chang}.

In order to gain insight into the QCD$_{2}$ origin of the
$\psi_{a}$ fields let us write the `mass term' in the multifermion
sector of ATM theory as \cite{prd}
\begin{eqnarray}
\label{massterm} {\bar{\psi}}_{a} \, e^{2i\varphi\,\gamma_5}\,
\psi^{a} = \psi^{\dagger \,a}_{L} \psi_{R\, a}\, e^{2i\varphi} +
\psi^{\dagger\, a}_{R} \psi_{L\, a}\, e^{-2i\varphi}.
\end{eqnarray}
The ATM mass term in the multifermion sector, Eq. (\ref{massterm}), must be compared to
the corresponding term in the bosonized QCD$_{2}$ in order to identify the fields related
to the flavor and color degrees of freedom, respectively.

 Therefore the total chiral invariant Lagragian including the kinetic terms for the quark
 fields becomes
\br
\label{atmcolor}
 \frac{1}{k}{\cal L} = \frac{1}{4} \partial_{\mu} \varphi \, \partial^{\mu} \varphi
+ i e_{\psi} \sum_{a} ( \bar{\psi}_{a} \gamma^{\mu} \partial_{\mu} \psi_{a}
- m_{\psi}\,  \bar{\psi}_{a} \,
e^{2i\varphi\,\gamma_5}\, \psi_{a}).
\er

 Although the QCD color degrees of freedom have a non-abelian symmetry we use abelian
 bosonization techniques in order to bosonize the fermions. This will be sufficient in
 order to reproduce various properties of the effective QCD$_{2}$ Lagrangian in this
 regime as presented in ref. \cite{ellis111}. So, let us introduce new boson field
 representations of the fermion bilinears as \cite{stone}

 \begin{eqnarray}
i: {\bar{\psi }}_{a}\gamma ^{\mu }\pa_{\mu} \psi_{a}: &=& -\frac{\a^2}{2\pi}
(\partial_{\nu }\phi_{a})^2\\
: {\bar{\psi }}_{a}(1\pm \gamma _{5})\psi_{a} : &=&
- \frac{c \mu}{\pi} :  e^{(\pm i \a \phi_{a})} :  ,\,\,\,\,\,\,\,
: {\bar{\psi }}_{a}\gamma ^{\mu }\psi_{a} : \,=\,-\frac{\a}{\pi}
\epsilon^{\mu \nu}\partial_{\nu }\phi_{a}
\label{boson1},
\end{eqnarray}
where $c=\frac{1}{2}\exp{(\gamma)}$, $\mu$ is an infrared regulator and $\a$ a real
parameter.

In order to compare to the related QCD Lagrangian describing the regime $m_{q} >> e_{c} $
\cite{ellis111}, which does not possess an exact chiral symmetry,  we must introduce
some chiral symmetry breaking terms in the Lagrangian (\ref{atmcolor}). The most direct
program for accomplishing this is simply to include certain chiral breaking terms in the
bosonized version of the ATM$+$color model given in (\ref{atmcolor}) in the form of
\br\nonumber
{\cal L}_{bos} &=& \frac{k}{4} \partial_{\mu} \varphi \, \partial^{\mu} \varphi
+  \sum_{a}^{N_{c}}\{ \frac{k \a^2 e_{\psi}}{2\pi} \partial_{\mu} \phi_{a} \,
\partial^{\mu} \phi_{a}
+ \frac{k m_{\psi} c \mu e_{\psi}}{\pi} \mbox{cos} (2\vp + 2\a \phi_{a}) - m_{a}
\phi_{a}^2 \} - \\
&& m_{0} \vp^2 - \sum_{a<b} m_{ab} \phi_{a} \phi_{b} - \sum_{a}m_{0a} \vp \phi_{a}
\label{bosonized}
\er
Notice that we have included certain bilinear terms in the scalar fields as the symmetry
breaking terms. Define the fields $\chi_{a}$ and $\Phi$ as
\begin{eqnarray}
\label{linear}
\chi_{a} \equiv \frac{2}{\beta} (\a \phi_{a} + \varphi);\,\,\,\,\, \Phi \equiv
\frac{1}{\sqrt{2 d}}(\vp - \frac{k \b e_{\psi}}{4\pi d}\sum_{a=1}^{N_{c}}{\chi_{a}}),
\,\,\,\, d \equiv \frac{k}{4}+\frac{k \b^2 N_{c} e_{\psi}}{2\pi}.
\end{eqnarray}

So, providing the relationships
\br
m_{a}&=&const,\,\,\,m_{ab}=m_{ba}=const.\, (a<b),\,\,\,m_{0a}=const,\,\,\,\, \forall a,\\
m_{ab} &=& \frac{m_{01}^2}{2m_{0}} - \frac{2 \d_{c}^2}{N_{c}} ,\,\,\,\,m_{a} =
\frac{m_{01}^2}{4m_{0}} + \frac{\d_{c}^2(N_{c}-1)}{N_{c}},\,\,\,\d_{c}\equiv
\frac{8 e_{c}^2 \a^2}{\b^2},\\
e_{\psi}&=&\frac{-\pi m_{01}}{4\a m_{0}}.
\er
the Lagrangian (\ref{bosonized}) becomes
\begin{eqnarray}
\nonumber
{\cal L}_{bos} &=& \frac{1}{2}  (\partial_{\mu} \Phi )^2 - \frac{2m_{0}}{k} (1+ \frac{2N_{c}
e_{\psi}}{\pi}) \Phi^2  +  \sum_{a} \{ \frac{1}{2} (\partial_{\mu }\chi_{a} )^{2}
+  2M^2(\cos\, \beta \, \chi_{a} ) \}-\\
&& \frac{ k^2 e_{\psi}^2 \b^2}{8 \pi^2 d} \sum_{a<b} \pa_{\mu}\chi_{a}\pa^{\mu}\chi_{b} - 2 e^2_{c} (\frac{N_{c}-1}{N_{c}}) \sum_{a}\chi_{a}^{2}+ \frac{4 e_{c}^2}{N_{c}} \sum_{a<b} \chi_{a} \chi_{b}
\label{qcd1}
\end{eqnarray}
where
\br \beta^2 = 4\pi,\,\,\,\,M^2=\frac{c\, m_{\psi}\mu k e_{\psi}}{2\pi},\,\,\,\,e_{\psi}=\frac{N_{c}-\frac{1}{2}k\pi\pm \sqrt{N_{c}^2+k\pi N_{c}-2\pi k+\frac{1}{4} \pi^2k^2}}{2k(N_{c}-1)}\label{epsi}.\er
The model (\ref{qcd1}) except the $a<b$ kinetic (the first term of the second line in (\ref{qcd1})) and the $\Phi$ terms reproduces the QCD$_{2}$ bosonized Lagrangian (in the regime $m_{q} >> e_{c}$) presented in \cite{ellis111}. Notice that the $\Phi$ field completely decouples from the rest of the fields. Moreover, in the opposite limit, i.e. the strong coupling regime and large $N$ limit we can verify that this field becomes a free massless field \cite{prd}.

Besides, the low-energy spectrum of
QCD$_{2}$ has ben studied by means of abelian
\cite{baluni} and non-abelian bosonizations \cite{gonzales, frishman}. In this limit the
 baryons
of QCD$_{2}$ are sine-Gordon solitons \cite{frishman}. In the large $N$ limit approach
(weak $e$ and small $m_{q}$) the SG theory also emerges \cite{salcedo}.

The question of confinement of the ``color'' degrees of freedom associated to the field
 $\psi$ in the ATM model by computing the string tension has been presented in \cite{prd}.
  In the {\sl fundamental} representation of the quarks it has been taken $\frac{m_{\psi}}{4\pi}= m_{q}$ and
 $k= 2N/\pi$. Then from (\ref{epsi}) one has $|e_{\psi}| = \frac{\pi}{2 \sqrt{N^2-N}}$.

Following \cite{ellis111} we define the baryon number as
\br
B= \frac{1}{\sqrt{\pi}}\sum_{k=1}^{N_{c}} \[\chi_{k}(+\infty) - \chi_{k}(-\infty)\]
\er

We seek for solutions of the field equations of motion in the static case
\br
\chi^{''}_{a} - 4 M^2 \sqrt{\pi}\, \mbox{sen} \sqrt{4\pi} \chi_{a} + \rho
\sum_{b>a}^{N_{c}} \chi^{''}_{b} - 4e_{c}^2 \(\chi_{a}-\frac{1}{N_{c}} \sum_{b=1}^{N_{c}}
\chi_{b}\) = 0,\label{eqstat}\\
\,\,\,\,\rho \equiv [N-1+4 \pi \sqrt{N(N-1)}]^{-1}.
\er

Depending on the boundary conditions for the fields $\chi_{k}(\pm \infty)$ we may have
certain nucleon states with $B= k N_{c},\, k \in \IZ$ (the baryon number is normalized
to be $N_{c}$ for the nucleon) or some quark solitons ($B = n,\,\,n=$integer non-multiple
of $N_{c}$). These type of solutions can be discussed by analyzing the field equations for
the static case \cite{ellis111}. In the low energy and strong coupling limit
($e_{c} >> m_{q}$) the nucleon states (baryons and  multibaryon) are described by
the generalized sine-Gordon solitons (see \cite{jhep5} and references therein), as can
be inferred from the form of the eqs. of motion (\ref{eqstat}) in this limit. Whereas,
the quark solitons exist for a sufficiently heavy
quark $m_q$, but have infinite energy, corresponding to a
string carrying the non-singlet color flux off to spatial
infinity, i.e. they exist in the opposite limit $m_{q} >> e_{c}$. These quark soliton
solutions disappear when
the meson mass parameter M is reduced to become
comparable to the gauge coupling strength $e_c$ (it has the dimension of mass in QCD2).
Let us search  for solutions such that $\chi_{a}(-\infty)=0$ for all $a$. Then,
at $x=+\infty$ one has
\br
\label{bc1}
4 M^2 \sqrt{\pi}\, \mbox{sen} \sqrt{4\pi} \chi_{a}(\infty) + 4e_{c}^2 \(\chi_{a}(\infty)-
\frac{1}{N_{c}} \sum_{b=1}^{N_{c}} \chi_{b}(\infty)\) = 0.
\er
The eq. (\ref{bc1}) becomes the same as the one presented in \cite{ellis111} describing
the
boundary condition at $x=+\infty $. If we assume $\chi_{a}(\infty)=\chi$ for all $a$,
one has that $\chi(\infty)=\frac{1}{2} \sqrt{\pi} n$, and $B=\frac{1}{2} n N_{c}$. But,
in order to have positive eigenvalues of the squared mass matrix $\frac{\pa^2 V}{\pa
{\chi_{a}}\pa {\chi_{b}}}$, we must have even $n$, and thus integer baryon number
$B=k N_{c}$ (baryons and multibaryons).

Following \cite{ellis111}, in the search for quark solitons let us first concentrate
on the case $N_{c}=2$. Then, eq. (\ref{bc1}) can be written as
\br
\label{infty1}
\mbox{sin} \sqrt{4\pi} \chi_{1}(\infty) &=& - \epsilon \sqrt{\pi} [\chi_{1}
(\infty)-\chi_{2}(\infty)],\\\mbox{sin} \sqrt{4\pi} \chi_{2}(\infty) &=& -
\epsilon \sqrt{\pi} [\chi_{2}(\infty)-\chi_{1}(\infty)] \label{infty2},
\er
where $\epsilon = \frac{e_{c}^2}{2\pi M^2}$. We may have non-baryonic solitons
with $B=n$ for odd values of $n$ (the quarks correspond to $n=1$). For $\epsilon << 1$
we can find a series of solutions with positive second derivative matrix. This solution
satisfies
\br
\chi_{2}(\infty) = -\chi_{2}(\infty) + n \sqrt{\pi},
\er
which together with (\ref{infty1})-(\ref{infty2}) provides
\br
\label{infty3}
\mbox{sin} \sqrt{4\pi} \chi_{1} = - \epsilon (\sqrt{4\pi} \chi_{1} - n \pi).
\er
Let us define $\xi =  \sqrt{4\pi} [\chi(\infty)-\frac{1}{2} n\sqrt{\pi}]$, then one has
that the solutions are
\br
\label{solutions}
\xi_{l} = \left\{ \begin{array}{ll}(\pi - \epsilon ) (2 l) & \mbox{for } n \,\,\mbox{even}\\
(\pi - \epsilon ) (2 l+1) & \mbox{for } n \,\,\mbox{odd}\end{array}\right.,
\er
in the limit where $(l \epsilon << \frac{1}{4})$.  The solutions (\ref{solutions}) correspond
 to excitations of ``colored" states and have infinite energy, with classical string tension
\br
T \approx \left\{ \begin{array}{ll} \pi\, e_{c}^2 \, (2 l)^2 & \mbox{for } n \,\,\mbox{even}\\
\pi\, e_{c}^2\,  (2 l+1)^2 & \mbox{for } n
\,\,\mbox{odd}\end{array}\right.. \er The single constituent quark
soliton corresponds to $n=1, (2l+1)=1$. Thus, we have shown that
QCD$_{2}$ has quark soliton solutions if the quark mass is sufficiently
large. These quark solitons disappear when the quark mass $m_{q}$
is reduced until the meson mass M  becomes comparable to the
dimensional gauge coupling strength $e_{c}$. The above picture can
be directly generalized for any $N_{c}$, see more details in
\cite{ellis111}.

\begin{figure}
\centering
\hspace{-2.4cm}\scalebox{0.4}{\includegraphics[angle=0]{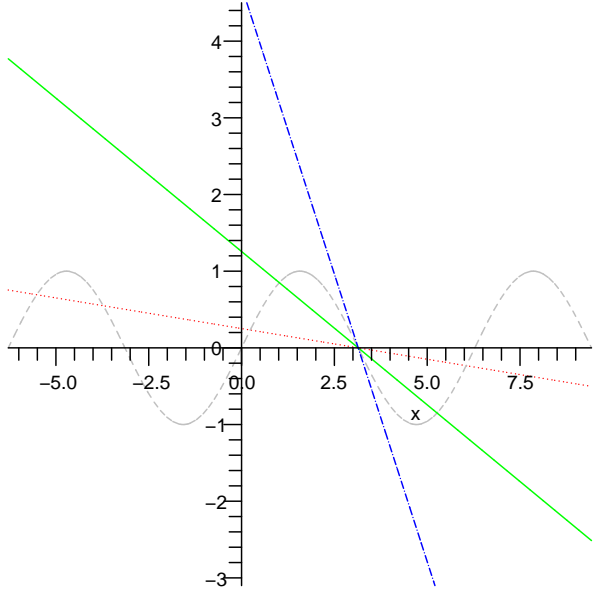}}
\parbox{5in}{\caption{The axis is defined as $x \equiv \sqrt{4\pi}
\chi_{1}(\infty)$. Comparison of the left-and right-hand sides of
the soliton eq. (\ref{infty3}), corresponding to a quark soliton
with $n= 1$, for $\epsilon = 0.08$ (dotted line) and
$\epsilon=0.4$ (solid line). There are no solutions for $\epsilon
> 1$ (dot-dashed line, drawn for $\epsilon = 1.5$).}}
\end{figure}

\section{Discussion}

The generalized sine-Gordon model GSG (\ref{eq1})-(\ref{eq2})
 provides a variety of solitons, kinks and bounce type solutions. The appearance
of the non-integrable double sine-Gordon model as a sub-model of
the GSG model suggests that this model is a non-integrable theory
for the arbitrary set of values of the parameter space. However, a
subset of values in parameter space determine some reduced
sub-models which are integrable, e.g. the sine-Gordon submodels of
subsections \ref{s11}, \ref{s22} and \ref{s33}.

In connection to the ATM spinors it was suggested that they are
confined inside the GSG solitons and kinks since the gauge fixing
procedure does not alter the $U(1)$ and topological currents
equivalence (\ref{equivalence}). Then, in order to observe the bag
model confinement mechanism it is not necessary to solve for the
spinor fields since it naturally arises from the currents
equivalence relation. In this way our model presents a bag model
like confinement mechanism as is expected in QCD.

The (generalized) massive Thirring model (GMT) is bosonized to the
GSG model \cite{epjc}, therefore, in view of the
 solitons and kinks found above as solutions of the GSG model we expect
 that the spectrum of the GMT model will contain $4$ solitons and their
 relevant anti-solitons, as well
as the kink and antikink excitations. The GMT Lagrangian describes
three flavor massive spinors with current-current interactions
among themselves. So, the total number of solitons which appear in
the bosonized sector suggests that the additional soliton
(fermion) is formed due to the interactions between the currents
in the GMT sector. However, in subsection \ref{s33} the soliton
masses $M_{3}$ and $M_{4}$ become the same for the case
$\mu_{1}=\mu_{2}$, consequently, for this case we have just three
solitons in the GSG spectrum, i.e., the ones with masses $M_{1}$,
$M_{2}$ (subsections \ref{s11}-\ref{s22} ) and $M_{3}=M_{4}$
(subsection \ref{s33}), which will correspond in this case to each
fermion flavor of the GMT model. Moreover, the $sl(3)$ GSG
model potential (\ref{potential}) has the same structure as the
effective Lagrangian of the massive Schwinger model with $N_{f}=3$
fermions,
 for a convenient value of the vacuum angle $\theta$. The
multiflavor Schwinger model resembles with four-dimensional QCD in
many respects (see e.g. \cite{hosotani} and references therein).

The $sl(n)$ ATM models may be relevant in the construction of
the low-energy effective theories of multiflavor QCD$_{2}$ with
the dynamical fermions in the fundamental and adjoint
representations. Notice that in these models the Noether and
topological currents and the generalized sine-Gordon/massive
Thirring models equivalences take place at the classical
\cite{jhep, annals} and quantum mechanical level \cite{epjc,
nucl1}.

The interest in baryons with 'exotic' quantum numbers has recently
been stimulated by various reports of baryons composed by four
quarks and an antiquark. The existence of these baryons cannot yet
be regarded as confirmed, however, reports of their existence have
stimulated new investigations about baryon structure (see e.g.
\cite{kabana} and references therein). Recently, the spectrum of
exotic baryons in QCD$_{2}$, with $SU(N_{f})$ flavor symmetry, has
been discussed providing strong support to the chiral-soliton
picture for the structure of normal and exotic baryons in four
dimensions \cite{jhep5, ellis}. The new puzzles in
non-perturbative QCD are related to systems with unequal quark
masses, so the QCD$_{2}$ calculation must take into account the
$SU(N_{f})$-breaking mass effects, i.e. for $N_{f}=3$ it must be
$m_{s} \neq m_{u, d}$. So, in view of our results above, the
properties of the GSG and the ATM theories may find some
applications in the study of mass splitting of baryons in
QCD$_{2}$ and the understanding of the internal structure of
baryons. Regarding this line of research, it has been shown that
the GSG model describes the low-energy spectrum of normal and
exotic baryons in QCD$2$ with unequal quark mass parameters
 \cite{jhep5}.

Finally, we have considered the quark soliton (qualiton) solutions of QCD$_{2}$ in the regime $e_{c}<< m_{q}$. In this context the role played by
the sl(2) ATM model is clarified. In fact, the qualitons arise if the color degrees of freedom are restored by coupling them to the Toda field and  convenient boundary conditions are imposed on the fields. So, we have shown that the sl(2) ATM model becomes a low-energy effective lagrangian describing the quark confinement mechanism in QCD$_{2}$. The equivalence between the Noether and topological currents (\ref{equiv}) is a crucial property of the ATM model in order to provide the confinement mechanism. This picture can be directly generalized to any number of flavors $N_{f}$ since a relationship analog to (\ref{equiv}) holds in that case, e.g. the $N_{f}=3$ case is presented in (\ref{equivalence}).

\vskip 1.0cm

 {\sl\ Acknowledgements}

HB thanks the Physics Department-UFMT (Cuiab\'a) and IMPA (Rio
de Janeiro) for hospitality. HLC
thanks FAPESP for
support.

\appendix

\section{The zero-curvature formulation of the ATM model}
\label{atmapp}

We summarize the  zero-curvature formulation of the $sl(3)$ ATM
model \cite{jmp, jhep, bueno}. Consider the zero curvature
condition \br \label{zeroc}
\partial_{+}A_{-}-\partial _{-}A_{+}+[A_{+},A_{-}]=0.
\er

The potentials take the form \br \label{aa1} A_{+}=-B F^{+}B^{-1},\quad
A_{-}=-\partial _{-}BB^{-1}+F^{-},\qquad \er with \br \label{aa2} F^{+}
\,=\,F_{1}^{+}+F_{2}^{+},\,\,\,\,\,\, F^{-} \,=\, F_{1}^{-}+F_{2}^{-}, \er where $B$
and $F_{i}^{\pm }$ contain the fields of the model \br \lab{F1}
F_{1}^{+}&=&\sqrt{im^{1}_{\psi}}\psi _{R}^{1}E_{\alpha
_{1}}^{0}+\sqrt{im^{2}_{\psi}}\psi _{R}^{2}E_{\alpha
_{2}}^{0}+\sqrt{im^{3}_{\psi}}\widetilde{\psi }_{R}^{3}E_{-\alpha _{3}}^{1},
\\
\lab{F2} F_{2}^{+}&=&\sqrt{im^{3}_{\psi}}\psi _{R}^{3}E_{\alpha
_{3}}^{0}+\sqrt{im^{1}_{\psi}} \widetilde{\psi
}_{R}^{1}E_{-\alpha_{1}}^{1}+\sqrt{im^{2}_{\psi}}\widetilde{\psi }
_{R}^{2}E_{-\alpha _{2}}^{1},
\\
\lab{3} F_{1}^{-}&=&\sqrt{im^{3}_{\psi}}\psi _{L}^{3}E_{\alpha
_{3}}^{-1}+\sqrt{im^{1}_{\psi}} \widetilde{\psi }_{L}^{1}E_{-\alpha
_{1}}^{0}+\sqrt{im^{2}_{\psi}}\widetilde{\psi } _{L}^{2}E_{-\alpha _{2}}^{0},
\\
\lab{F4} F_{2}^{-}&=&\sqrt{im^{1}_{\psi}}\psi _{L}^{1}E_{\alpha
_{1}}^{-1}+\sqrt{im^{2}_{\psi}}\psi _{L}^{2}E_{\alpha
_{2}}^{-1}+\sqrt{im^{3}_{\psi}}\widetilde{\psi }
_{L}^{3}E_{-\alpha _{3}}^{0},\\
B&=&e^{i\th_{1} H^{0}_{1}+i\th_{2} H^{0}_{2} }\,e^{\widetilde{\nu }C}\,e^{\eta
  Q_{ppal}}\equiv b\, e^{\widetilde{\nu }C}\,e^{\eta
  Q_{ppal}}. \lab{equn1}
\er

$E_{\alpha _{i}}^{n},H^{n}_{1},H^{n}_{2}$ and  $C$ ($i=1,2,3; \, n=0,\pm 1$) are some
generators of $sl(3)^{(1)}$; $Q_{ppal}$ being the principal gradation operator. The
commutation relations for an affine Lie algebra in the Chevalley basis are \br
&&\left[ \emph{H}_a^m,\emph{H}_b^n\right] =mC\frac{2}{\alpha_{a}^2}K_{a b}\delta _{m+n,0}  \lab{a7}\\
&&\left[ \emph{H}_a^m,E_{\pm \alpha}^n\right] = \pm K_{\alpha a}E_{\pm \alpha}^{m+n}
\lab{a8}\\
&&\left[ E_\alpha ^m,E_{-\alpha }^n\right] =\sum_{a=1}^rl_a^\alpha
\emph{H}_a^{m+n}+\frac 2{\alpha ^2}mC\delta _{m+n,0}  \lab{a9}
\\
&&\left[ E_\alpha ^m,E_\beta ^n\right] = \varepsilon (\alpha ,\beta )E_{\alpha +\beta
}^{m+n};\qquad \mbox{if }\alpha +\beta \mbox{ is a root \qquad }  \lab{a10}
\\
&&\left[ D,E_\alpha ^n\right] =nE_\alpha ^n,\qquad \left[ D,\emph{H}%
_a^n\right] =n\emph{H}_a^n.  \lab{a12} \er where $K_{\alpha
a}=2\a.\a_{a}/\a_{a}^2=n_{b}^{\a}K_{ba}$, with $n_{a}^{\a}$ and $l_a^\alpha$ being
the integers in the expansions $\a=n_{a}^{\a}\a_{a}$ and
$\a/\a^2=l_a^\alpha\a_{a}/\a_{a}^2$, and $\varepsilon (\alpha ,\beta )$ the relevant
structure constants.

Take $K_{11}=K_{22}=2$ and $K_{12}=K_{21}=-1$ as the Cartan matrix elements of the
simple Lie algebra $sl(3)$. Denoting by $\a_{1}$ and $\a_{2}$ the simple roots and
the highest one by $\psi (=\a_{1}+\a_{2})$, one has $l_{a}^{\psi}=1(a=1,2)$, and
$K_{\psi 1}=K_{\psi 2}=1$. Take $\varepsilon (\alpha ,\beta )=-\varepsilon (-\alpha
,-\beta ),\,\, \varepsilon_{1,2}\equiv \varepsilon (\alpha_{1} ,\a_{2})=1,\,\,
\varepsilon_{-1,3}\equiv \varepsilon(-\alpha_{1} ,\psi )=1\,\, \mbox{and}\,
\,\,\varepsilon_{-2,3}\equiv \varepsilon (-\alpha_{2} ,\psi)=-1$.

One has $Q_{ppal} \equiv \sum_{a=1}^{2}  {\bf s}_{a}\l^{v}_{a}.H +
3 D$, where $\l^{v}_{a}$ are the fundamental co-weights of
$sl(3)$, and the principal gradation vector is ${\bf s}=(1,1,1)$
\cite{kac}.

\end{document}